\documentclass[pra,showpacs,twocolumn,superscriptaddress,floatfix]{revtex4}

\newcommand{\avs}[1]{\left<#1\right>}

\newcommand{\eff}{{\text{eff}}}

\usepackage{amsfonts}
\usepackage{amsmath}
\usepackage{graphicx}
\usepackage{amssymb}
\usepackage{color}
\usepackage{subfigure}
\usepackage[perpage]{footmisc}
\usepackage{mathtools}

\begin{document}
\title{Mechanical back-reaction effect of the dynamical Casimir emission}

\author{Salvatore Butera}
\author{Iacopo Carusotto}
\affiliation{INO-CNR BEC Center and Dipartimento di Fisica, Universit\`a di Trento, I-38123 Povo, Italy}
\begin{abstract}
We consider an optical cavity enclosed by a freely moving mirror attached to a spring and we study the quantum friction effect exerted by the dynamical Casimir emission on the mechanical motion of the mirror. Observable signatures of this simplest example of back-reaction effect are studied in both the ring-down oscillations of the mirror motion and in its steady-state motion under a monochromatic force. Analytical expressions are found in simple yet relevant cases and compared to complete numerical solution of the master equation. A circuit-QED device allowing for experimental observation of the effect with state-of-the-art technology is proposed and theoretically characterized.
\end{abstract}
\maketitle

\makeatletter
\let\toc@pre\relax
\let\toc@post\relax
\makeatother 

\section{Introduction\label{sec:Intro}}
Quantum field theories on curved spacetime represent a first step towards the long-sought general quantum theory unifying all the known fundamental interactions, including gravity. Analogously to the semi-classical theory of light-matter interaction, where light is considered as a classical field governed by the Maxwell equations, while the internal dynamics of the atoms is quantized, these field theories describe the spacetime as a classical background ruled by Einstein's theory of general relativity, while matter is described as quantized fields. 

The origin of this semiclassical theory of the interaction between gravity and matter dates back to the early '70s with the seminal works by Parker \cite{Parker-PartCr-I,Parker-PartCr-II} and Hawking \cite{Hawking1975}, who extended the quantum field theory of elementary particles from the (flat) Minkowski spacetime of special relativity to more general curved spacetimes that are solutions of the nonlinear Einstein gravitational field equations. This first attempt to consider the effects of gravity on quantum fields opened the door to a plethora of new intriguing effects, that were unexpected in standard quantum field theories on Minkowskian spacetime. Particle creation in expanding universes, and the evaporation of black holes (BH) in the form of a thermal radiation were the first and still among the most prominent predictions of this new theory.
On the basis of these first outcomes, it was soon realized that particle creation out of the quantum vacuum state is a very general effect, that, e.g., takes place whenever the background spacetime is non-stationary and/or displays a horizon. A similar phenomenon was predicted by Fulling and Davies \cite{Fulling-Davies,Davies-Fulling}, who anticipated that photons are created out of the vacuum state when a non-uniformly accelerating boundary conditions (describing, e.g., a moving mirror) is imposed to the field itself. In the literature, this effect is generally known as dynamical Casimir effect (DCE) \cite{Schwinger-DCE,Yablonovitch-DCE,Dodonov-DCE}, a term which emphasizes its origin from the same zero-point fluctuations of quantum fields that give rise to the Casimir force between neutral objects \cite{Milton-book,Mostepanenko-book,Dalvit-book,Milonni-book,Plunien-casimir}.

The standard derivation of the Hawking radiation, cosmological particle creation, and DCE is based on the semi-classical assumption according to which a quantum field lives on a fixed back-ground, whose geometry is not affected by the dynamics of the field itself. This simplifying hypothesis leads to theories that, strictly speaking, are not self-consistent and violate basic physical principles such as energy conservation and unitary evolution. The formulation of a fully consistent theory need to take into account the back-reaction of the quantum field on the back-ground spacetime, or on the boundary conditions in the DCE case. 

The problem of the back-reaction by particles created in non-stationary spacetimes has a long history. In the context of cosmological particle creation, such a back-reaction manifests as a damping of the expansion of the universe \cite{Zeldovich-BR-Cosm-1972,Hu-BR-Cosm-1973,Parker-BR-Cosm-1973,Hu-BR-Cosm-1973-I,Hu-BR-Cosm-1973-II,Hartle-BR-Cosm-1977,Hu-BR-Cosm-1978,Hu-BR-Cosm-1979-I,Hu-BR-Cosm-1979-II,Hu-BR-Cosm-1979-III,Shaw-BR-Cosm1999}. In the case of Hawking radiation, one instead expects from purely thermodynamic considerations that the horizon of a black-hole must shrink as a consequence of the emitted particles \cite{Bekenstein-BH-Entr-1972,Hawking-BH-Entr-1973,Bekenstein-BH-Entr-1973,Page-BH-Therm-2005,Wald-BH-Therm-2001}. In the DCE case, the particles created out of the vacuum provide a friction force on the moving mirror \cite{Oku-BR-DCE-1979,Xuereb-BR-2009,Xuereb-BR-2012,Hu-MOF-2013,Hu-MOF-2015}. Despite the efforts devoted to this topic, most works so far assume that the background interacts with expectation values of quantum field observables such as the stress-energy tensor. A clear picture of the exact quantum dynamics of these processes is therefore still unknown and calls for further investigations: a full understanding of the back-reaction problem will be a crucial brick in the perspective of building a fully consistent theory of the gravitational interaction between spacetime and quantum fields and will likely provide an answer to fundamental open questions such as the long-time fate of a black-hole and of the information it has swallowed during its lifetime. 

A first step in this direction is the mechanical back-reaction of DCE particles onto the motion of a moving mirror. In its simplest formulation, this problem can be formulated in terms of a single-mode cavity enclosed by mechanically movable mirrors, coupled to each other via the radiation pressure effect. This simplified geometry enormously reduces the complexity of the problem and allows for a good understanding of the underlying physics in terms of quanta of mechanical motion being converted into pairs of photons, which then exert a back-reaction effect on the mirror in the form of a mechanical friction.


%
While recent advances in the miniaturization technologies have led to a variety of strictly opto-mechanical effects including the cooling of the mechanical oscillator to its quantum mechanical ground state \citep{Aspelmeyer_RMP}, the conversion of quantum fluctuations in the cavity vacuum state into real photons via the DCE has so far escaped experimental observation \cite{Braggio-MIR-2008,Dodonov-DCE}. The main reason for that is the wide separation in frequency of the (high-frequency) optical and (low-frequency) mechanical modes, that hinders fulfilment of the DCE resonance condition $\omega_b\approx 2 \omega_a$ between the cavity and mechanical frequencies $\omega_{a,b}$ and thus dramatically suppresses the intensity of the DCE emission. Higher-harmonic couplings may be exploited to release the resonant condition to $n_b\omega_b=2\omega_a$ ($n_b$ being an integer) \cite{Lambrecht-2005}, but the efficiency of the resulting DCE remains quite low. Very recently, a dramatically reinforced efficiency in strongly nonlinear ultra-strong light-matter coupling regimes was predicted in \cite{Savasta-PRX-2018}, where a first investigation of the back-reaction effect of DCE onto the mirror was also reported.

Since Unruh's original proposal of analog black holes \cite{Unruh-Analog-1981}, the general concept of analog system turned out to be a fruitful framework where to study physical phenomena whose experimental investigation is out of current technological capabilities~\cite{Barcelo-2011}. The basic idea is to look for an experimentally controllable system, whose dynamics is governed by the same equations of motion of the system of interest, but in a completely different physical context and energy scales. In the original proposal \cite{Unruh-Analog-1981}, the goal was to experimentally verify Hawking's prediction of black hole evaporation using acoustic waves in non-uniformly flowing fluids. The first experimental success of the analog model idea was the demonstration of DCE in a circuit-QED context \cite{Wilson-DCE-Analog-2011}. A superconducting quantum interference device (SQUID)  was used to impose a magnetically tunable boundary condition to the electromagnetic field in a coplanar waveguide, analogous to an effective mirror whose spatial position is controlled by the applied magnetic field. When the position of this (analog) mirror is made to oscillate in space via a suitable modulation of the magnetic field threaded through the SQUID, a sizable DCE emission into the waveguide was observed, spectrally centered at half the modulation frequency and displaying peculiar the same quantum optical properties expected in the DCE \cite{Wilson-DCE-Analog-2011}. Since no mechanically moving element was present, the experiment belongs to the class of analog models. However, its quantum evolution equations are identical to the one of the standard DCE effect. Whereas we focus our attention on the experimental set-up in~\cite{Wilson-DCE-Analog-2011} , similar ideas can be developed for the related DCE experiment that was published shortly after in~\cite{Lahteenmaki-DCE-Analog-2013}.

A first theoretical study of back-reaction effects in all-optical analog models of DCE was reported in \cite{Iacopo-DCE-OptAn-2012}. While a strong and experimentally observable signal of back-reaction was anticipated there, the proposed device required a quite complex optical set-up and the connection to the general physics of DCE remained non-trivial. It is therefore important to devise configurations that allow for a direct insight into the basic physics of back-reaction and are promising in view of experimental realization using state-of-the-art technology. Here we consider a direct extension of the device used in \cite{Wilson-DCE-Analog-2011}, where the SQUID is no longer driven by a classical, pre-determined external field $B(t)$, but is magnetically coupled to an external $LC$ resonator that plays the role of the harmonically moving mirror: as a key novel feature of this work, we treat the $LC$ resonator as an independent dynamical degree of freedom and we show that the equations describing the coupled dynamics of the $LC$ and the waveguide are equivalent to the ones for a perfect, harmonically trapped mirror interacting with a quantum electromagnetic field via its radiation pressure \cite{Law-MirFieldInt-1995}. Quantitative estimates for the strength of the analog optomechanical coupling between the effective moving mirror and the cavity are given, as well as for the back-reaction effect.

The article is organized as follows. We start by introducing in Sec.~\ref{sec:System} the physical system at hand and by revising the fundamental concepts of the opto-mechanical interaction between the mechanical and electromagnetic degrees of freedom. In Sec.~\ref{subsec:ParOsc} we then briefly review the mean-field theory of the system dynamics, which models the evolution of the system in the classical limit. In order to describe strictly quantum effects such as particle creation from DCE and the back-reaction effects, a more sophisticated theory going beyond the mean-field approximation is developed in Sec.~\ref{subsec:QuasiMean}.
The key results of our work are presented in Sec.~\ref{sec:VacuumFriction}. The back-reaction is first investigated in Sec.~\ref{subsec:Free} for the case of an initially displaced mirror that performs free oscillations while interacting with the cavity mode. For relatively weak opto-mechanical coupling strength, the back-reaction results in a reinforced damping of the mechanical oscillation. For coupling strengths stronger than the loss rate, the back-reaction results instead in a periodic and reversible exchange of energy between the mirror and the field. 
In Sec.~\ref{subsec:Driven} we then study the novel configuration where the mirror is mechanically driven by a monochromatic external force: for a weak opto-mechanical coupling, the back-reaction effect is visible as a broadened lineshape for the resonant mechanical response of the mirror. For stronger couplings, we anticipate a splitting of the resonant response into a pair of Rabi-split peaks as well as a number of other nonlinear features. We conclude the paper with a discussion in Sec.~\ref{sec:Circuit} of a possible implementation of this physics in an analog model that directly extends the set-up used for the first experimental observation of the DCE \cite{Wilson-DCE-Analog-2011}. Promising estimates for the strength of the back-reaction effect are found using  parameters of state-of-the-art devices. Conclusions and future perspectives are finally given in Sec.~\ref{sec:Conclusions}.

\section{The system\label{sec:System}}
\begin{figure*}[t]
\centering%
\includegraphics[width=0.7\linewidth]{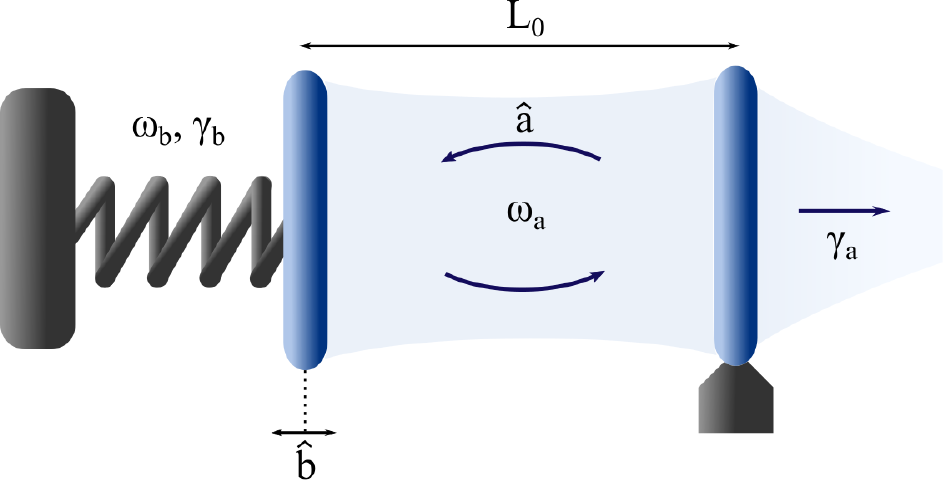}
\caption{Illustrative representation of a generic optomechanical system. One of the cavity mirrors is allowed to harmonically oscillate around the position of equilibrium and is opto-mechanically coupled to the cavity mode via the radiation pressure.}
\label{fig:sketch}
\end{figure*}

We consider a system (see the sketch in Fig.\ref{fig:sketch}) composed by an optical cavity terminated on one side by a mechanically moving mirror of mass $m_b$, confined around its equilibrium position by a harmonic potential of characteristic angular frequency $\omega_b$. We restrict the dynamics of the field to a single relevant mode of the optical cavity and we indicate with $\omega_a$ the frequency of the cavity mode when the mirror is at its equilibrium position and the cavity has length $L_0$.

Defining by $\hat{a}/\hat{a}^\dag$ and $\hat{b}/\hat{b}^\dag$ the annihilation/creation operators for the field and the mechanical oscillator respectively, the Hamiltonian $\hat{H}_0$ for the non interacting system takes the simple form
\begin{equation}
	\hat{H}_0=\hbar\omega_a \hat{a}^\dag \hat{a} +\hbar\omega_b \hat{b}^\dag \hat{b}.
\label{H0}
\end{equation}
The mirror and the field interact with each other via the radiation pressure, defined in terms of a pressure operator $\hat{F}_0$ which depends quadratically on the field \cite{Law-MirFieldInt-1995},
\begin{equation}
	\hat{F}_0=\frac{\hbar\omega_a}{2L_0}\,\left(\hat{a}+\hat{a}^\dag\right)^2
\label{F0}.
\end{equation}
In terms of the displacement operator for the mirror around its equilibrium position $\hat{x}=x_{\text{ZPF}}(\hat{b}+\hat{b}^\dag)$, where $x_{\text{ZPF}}=\left(\hbar/2m\omega_b\right)^{1/2}$ is the amplitude of the mechanical zero-point fluctuations, the opto-mechanical pressure interaction is described by the Hamiltonian \cite{Aspelmeyer_RMP}
\begin{equation}
	\hat{H}_{\text{int}}=\hat{x}\hat{F}_0=\hbar\omega_c\left(\hat{a}+\hat{a}^\dag\right)^2\left(\hat{b}+\hat{b}^\dag\right).
\label{Hint}
\end{equation}
where the strength of the opto-mechanical coupling between the mechanical and electromagnetic degrees of freedom is quantified by the effective interaction frequency 
\begin{equation}
\omega_c=\frac{\omega_a x_{\text{ZPF}}}{2L_0}=\left(\frac{\hbar}{8m_b\omega_b} \right)^{1/2}\left(\frac{\omega_a}{L_0}\right) 
\label{omegaC}
\end{equation}

For the sake of simplicity, we assume from now on that the system in a regime where the opto-mechanical coupling is much weaker than the natural oscillation frequencies of the both the cavity and the mechanical mirror, $\omega_c/\omega_{a/b}\ll 1$. Such an assumption does not represent a significant limitation for our purposes, but allows to neglect extra effects such as the dressing of the mirror by virtual photons and the consequent modification of the ground state of the interacting system \cite{Giulio-PRL-2013}. More specifically, under this condition the effects of anti-resonant terms of the Hamiltonian in Eq.~\eqref{Hint} like $\hat{a}^\dag\hat{a}\hat{b}$, $\hat{a}^2b$ can be neglected, as they are responsible for a minor correction to the energy levels of the system \cite{CohenTannoudji-AtomPhot}. The opto-mechanical coupling is thus modelled by the resonant terms $\hat{a}^2\hat{b}^\dag$, $\left(\hat{a}^\dag\right)^2\hat{b}$ only, which describe the creation (resp. annihilation) of mechanical excitations in the mirror and the simultaneous annihilation (resp. creation) of a  pair of photons. This is the physical mechanism responsible for the DCE, and thus for the exchange of energy between the mirror and the field and, in the final instance, for the appearance of friction in the mechanical motion of the mirror. A more general numerical approach that includes the ultra-strong coupling limit $\omega_c/\omega_{a/b}\gtrsim 1$ and the effects of the anti-resonant terms was recently pursued in \cite{Savasta-PRX-2018}.

We consider that both the mirror and the cavity mode are coupled to external degrees of freedom. In particular, we assume that the mirror is mechanically  driven by an external coherent force of amplitude $F(t)$, which can be modelled by means of additional time-dependent terms in the Hamiltonian
\begin{equation}
	\hat{H}_{\text{drive}}=-\hbar\left(\hat{b} F^*(t)+\hat{b}^\dag F(t)\right).
\label{Drive}
\end{equation}
Summing up all terms, the coherent dynamics of the system is modelled by the total Hamiltonian 
\begin{equation}
\begin{split}
	\hat{H}&=\hat{H}_0+\hat{H}_{\text{int}}+\hat{H}_{\text{drive}}\\
	&=\hbar\omega_a\hat{a}^\dag \hat{a}+\hbar\omega_b\hat{b}^\dag \hat{b}+\hbar\omega_c\left(\hat{b}^\dag \hat{a}^2+\hat{b}\left(\hat{a}^\dag\right)^2\right)\\
	&\quad-\hbar\left(\hat{b} F^*(t)+\hat{b}^\dag F(t)\right).
\end{split}
\label{Hs}
\end{equation}
On top of this, we take into account losses in the system by coupling the optical field to an external radiative and/or non-radiative baths and by including mechanical dissipation damping the mirror motion. Both these effects are included at the level of the master equation, so that the time evolution of the density matrix $\hat{\rho}$ of the interacting mirror-field system has the form
\begin{equation}
	\frac{d\hat{\rho}}{dt}=\frac{1}{i\hbar}\left[\hat{H},\hat{\rho}\right]+\mathcal{L}_{\hat{a}}[\hat{\rho}]+\mathcal{L}_{\hat{b}}[\hat{\rho}],
	\label{MasterEq}
\end{equation}
in terms of the Lindblad superoperators 
\begin{equation}
\mathcal{L}_{\hat{o}}\equiv\left(\gamma_o/2\right)\left(2\hat{o}\hat{\rho} \hat{o}^\dag-\hat{o}^\dag \hat{o} \hat{\rho}-\hat{\rho} \hat{o}^\dag \hat{o}\right) 
\end{equation}
describing cavity and mechanical losses with $\hat{o}=\hat{a},\hat{b}$, respectively. 
Equation of motion for expectation value of generic observables $\hat{O}$ can finally be obtained from the master equation,
\begin{multline}
	\frac{d\left<\mathcal{O}\right>}{dt}=\frac{1}{i\hbar}\text{Tr}_S\left\{\left[\hat{\mathcal{O}},\hat{H}\right]\hat{\rho}\right\}\\
	+\sum_{\hat{o}=\hat{a},\hat{b}}\frac{\gamma_o}{2}\left(\text{Tr}\left\{\left[\hat{o}^\dag,\hat{\mathcal{O}}\right]\hat{o}\,\hat{\rho}\right\}+\text{Tr}\left\{\hat{o}^\dag\left[\hat{\mathcal{O}},\hat{o}\right]\hat{\rho}\right\}\right).
\label{EvolOperators}
\end{multline}

In the next sections we are going to develop a formalism to obtain explicit results for the quantum system dynamics, which is able to go beyond the mean-field approximation and take into account the quantum fluctuations of the field at the simplest level. Based on this, we will provide a quantitative estimate for the friction due to the emission of dynamical Casimir pairs, and we will compare the analytical results with the full numerical solution of the master equation in Eq.~\eqref{MasterEq}.

\section{Theoretical models\label{sec:Models}}
\subsection{Mean-field theory of the parametric oscillator\label{subsec:ParOsc}}
The cubic nature of the Hamiltonian in Eq.~\eqref{Hs} makes the solution of the interacting field-mirror problem far from trivial. Simplifying hypothesis are thus needed, in order to derive approximate solutions which are able to capture at least some of the most significant properties of the system. In the semi-classical limit, one restricts the study to the average value of the amplitude of the field and mirror oscillations and replaces the $\hat{a}$ and $\hat{b}$ operators with the corresponding classical variables $a\equiv \langle \hat{a}\rangle$ and $b\equiv \langle\hat{b}\rangle$. The equations of motion for such \emph{mean-field} components can be derived from the master equation. 

Assuming that the drive is monochromatic $F(t)=F_0\,e^{-i\omega t}$ with given amplitude $F_0$ and frequency $\omega$, we can move to the frame rotating with the angular frequency $\omega$ of the drive. Within the rotating frame, the annihilation operators transform to $\hat{a}\rightarrow \hat{a}\,e^{-i\omega t/2}$ and $\hat{b}\rightarrow \hat{b}\,e^{-i\omega t}$, so that the equation of motion for the expectation values get the autonomous form 
\begin{align}
\frac{da}{dt}&=-\left(\frac{\gamma_a}{2}-i\Delta_a\right) a-2i\omega_c a^* b,
\label{MeanFieldA}\\
\frac{db}{dt}&=-\left(\frac{\gamma_b}{2}-i\Delta_b\right) b-i\omega_c a^2+i F_0,
\label{MeanFieldB}
\end{align}
where we defined the detuning $\Delta_a\equiv\omega/2-\omega_a$ and $\Delta_b\equiv\omega-\omega_b$. Given the dissipative form of Eqs.~\eqref{MeanFieldA} and \eqref{MeanFieldB}, a steady-state solution can be derived by setting the time derivatives to zero. 
These equations have the simplest form in the fully resonant case where the drive is resonant with the mirror frequency ($\omega=\omega_b$) and this latter is in resonance with twice the optical frequency ($\omega_b=2\omega_a$).

In these conditions the system exhibits a sort of phase transition at the threshold value $F_0=F_{0}^{\text{th}}\equiv \gamma_a \gamma_b/8\omega_c$ of the drive amplitude, at which the solution
\begin{align}
 a_B&=0,
\label{ABelowTr}\\
 b_B&=\frac{2i}{\gamma_b} F_0,
\label{BelowTr}
\end{align}
that is stable below the threshold $F_0<F_{0}^{\text{th}}$, becomes dynamically unstable. Above threshold, the system spontaneously break a $Z_2$ symmetry and has the choice to migrate towards two possible different branches, characterized by the same mirror amplitude but equal and opposite values of the field amplitude,
\begin{align}
 a_A^\pm&=\pm\left(\frac{F_0-F_0^{\text{th}}}{\omega_c}\right)^{1/2},
\label{AAboveTr}\\
 b_A&=i\frac{\gamma_a}{4\omega_c}.
\label{BAboveTr}
\end{align}
The parametric oscillator threshold at $F_0^{\text{th}}$ thus separates two qualitatively different regimes of the system. \emph{Below} the threshold, the classical component of the cavity field is zero, while the average amplitude of the mechanical oscillations increase linearly with the strength of the applied drive. As we shall see shortly, in this regime the quantum fluctuations of the field play a major rule in determining the quantum state of the cavity field. 
Conversely, \emph{above} the threshold, the expectation value of the field is finite and the mirror amplitude saturates to a finite value. In this case the system behaves to a good approximation classically, with the quantum fluctuations accounting only for small corrections to the mean-field dynamics. For later convenience we define $\gamma_0^2\equiv\gamma_a\gamma_b/2$, so that $F_0^{\text{th}}=\gamma_0^2/4\omega_c$.

While this classical model is typically able to reproduce the general trend of the steady-state field expectation values, it is not able to capture strictly quantum effect, such as the parametric amplification of vacuum fluctuations of the electromagnetic field and, in turn, the back-reaction of the dynamical Casimir photons onto the mechanical degrees-of-freedom. This can be directly seen from the mean-field steady-state below threshold found above, which contains no cavity excitation $a_B=0$. Generalization of the steady-state solutions (\ref{ABelowTr}-\ref{BelowTr}) to general values of $\Delta_b$ further shows that the response function of the oscillator to the external drive has the form of a Lorentzian function with central frequency $\omega_b$, and a linewidth equal to the damping rate $\gamma_b$ of the bare mechanical oscillator, 
\begin{equation}
	|b(\omega)|^2=\frac{|F_0|^2}{\Delta_b^2+(\gamma_b/2)^2}.
\label{BareRespFunc}
\end{equation}
The absence of any Casimir emission and any back-reaction effect shows that, in order to understand the physics of these effects it is necessary to go beyond the mean-field approximation and include quantum fluctuations in the model. This will be the objective of the following sections.

\subsection{Beyond mean-field\label{subsec:QuasiMean}}
In order to go beyond the mean-field theory, we first note that, because of the symmetry properties of the Hamiltonian in Eq.~\eqref{Hs}, the expectation value of any correlator containing an odd number of cavity field annihilation and creator operators $\hat{a},\hat{a}^\dagger$ does not change in time under the Hamiltonian evolution and remains strictly zero in the steady-state. The fundamental dynamical quantities for the field are thus given by the quadratic operators $\hat{q}\equiv \hat{a}^2$ and $\hat{n}_a\equiv \hat{a}^\dag \hat{a}$.

On this basis, a simple description of the quantum dynamics of the system can be formulated in terms of the time evolution of the expectation values of the amplitudes $b\equiv\avs{\hat{b}}$ and $q\equiv\avs{\hat{q}}$ for the mirror and the cavity field respectively, and of the number of photons in the cavity $n_a\equiv\avs{\hat{n}_a}$. Working again in the frame rotating at the drive frequency $\omega$, we can describe the system by the set of three equations
\begin{align}
	\frac{db}{dt}&=-\left(\frac{\gamma_b}{2}-i\Delta_b\right) b-i\omega_c q+i F_0,
\label{EqMotForc-1}\\
	\frac{dn_a}{dt}&=-\gamma_a n_a-2i\omega_c\left<q^\dag b\right>_S+2i\omega_c\left<q b^\dag\right>_S,
\label{EqMotForc-2}\\
		\frac{dq}{dt}&=-\left(\gamma_a-i\Delta_q\right)q-4i\omega_c\left<n_a b\right>_S-2i\omega_c b,
\label{EqMotForc-3}
\end{align}
where we defined the detuning $\Delta_q\equiv\omega-2\omega_a$. The Eqs.~(\ref{EqMotForc-1}-\ref{EqMotForc-3}) reveal how presence of cubic terms in the Hamiltonian Eq.~\eqref{Hs} leads to an infinite hierarchy of correlators, that need to be suitably truncated in order to obtain a solution to the problem. This effectively means neglecting the correlation of higher order between the mirror and the field, and attention must be paid to the conditions under which this approximation is justified. To this aim we identify three different regimes. 

i) In the limit of a weak drive $F_0\ll F_0^\text{th}$, the correlators involving products of two $\hat{b}$, $\hat{q}$ and $\hat{n}_a$ operators can be safely neglected as they represent higher order terms in the infinitesimal quantities $q$, $b$ and $n_a$. From now on, this regime will be called \emph{linear regime}, since in this case the Eqs.~(\ref{EqMotForc-1}-\ref{EqMotForc-3}) reduce to a set of three linear equations. In spite of its simplicity, this linear model is able to account for the quantum fluctuations responsible for the DCE emission and, then, for the back-reaction effect.

ii) In the opposite limit of a strong drive $F_0\gg F_0^\text{th}$, the system is in the parametric oscillator limit. As mentioned in the previous section, in this regime the system behaves in an approximately classical way, and the non-factorisable component (that is the \emph{cumulant} in the language of statistics) in the correlations between the field and the mirror can be neglected, with the latter factorizing as $\langle a^2 b^\dag\rangle=qb^*$ and $\langle a^\dag a b\rangle=n_a b$. The equations of motion (\ref{EqMotForc-1}-\ref{EqMotForc-3}) then reduce to the closed nonlinear set
\begin{align}
	\frac{d b}{dt}&=-\left(\frac{\gamma_b}{2}-i\Delta_b\right) b-i\omega_c q+i F_0,
\label{EqMotSoft-1}\\
	\frac{d n_a}{dt}&=-\gamma_a n_a-2i\omega_c q^* b+2i\omega_c q b^*,
\label{EqMotSoft-2}\\
		\frac{d q}{dt}&=-\left(\gamma_a-i\Delta_q\right) q-4i\omega_c n_a b-2i\omega_c b.
\label{EqMotSoft-3}
\end{align}

iii) In the region of parameters between these two limits, that is for $F_0\sim F_0^{\text{th}}$, quantum fluctuations plays a relevant role and the non-trivial higher order correlations between the field and the mirror need to be fully taken into account to properly describe the properties of the system.

For the purpose of this article, we note that the best conditions for the investigation of the back-reaction effects from the DCE photons are met in the linear regime. In this case, all the the key features of the DCE mechanism are kept into play, with the advantage of being able to neglect all the complex nonlinear effects arising from the radiation pressure coupling of the field with the mechanical oscillator. As we will see in the next sections, this simplifies very much the analysis, and closed expressions for the quantities of interest can be obtained by analytical means.

\section{Vacuum-induced friction \label{sec:VacuumFriction}}

\subsection{Free evolution\label{subsec:Free}}

\begin{figure*}[htbp]
\centering%
\subfigure [\label{fig:FreeB2G5}]
{\includegraphics[width=0.45\linewidth]{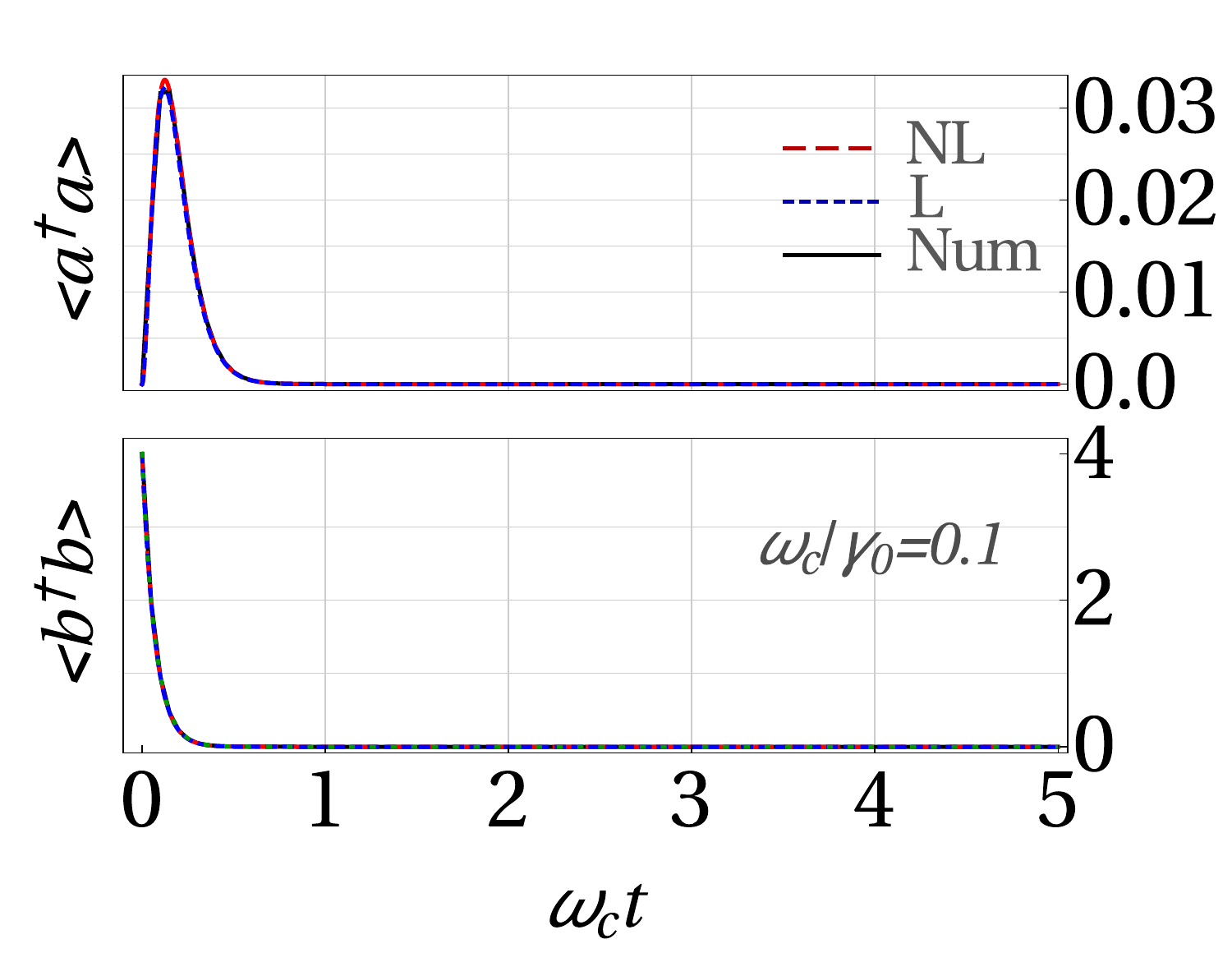}}\qquad
\subfigure [\label{fig:FreeB2G2}]
{\includegraphics[width=0.45\linewidth]{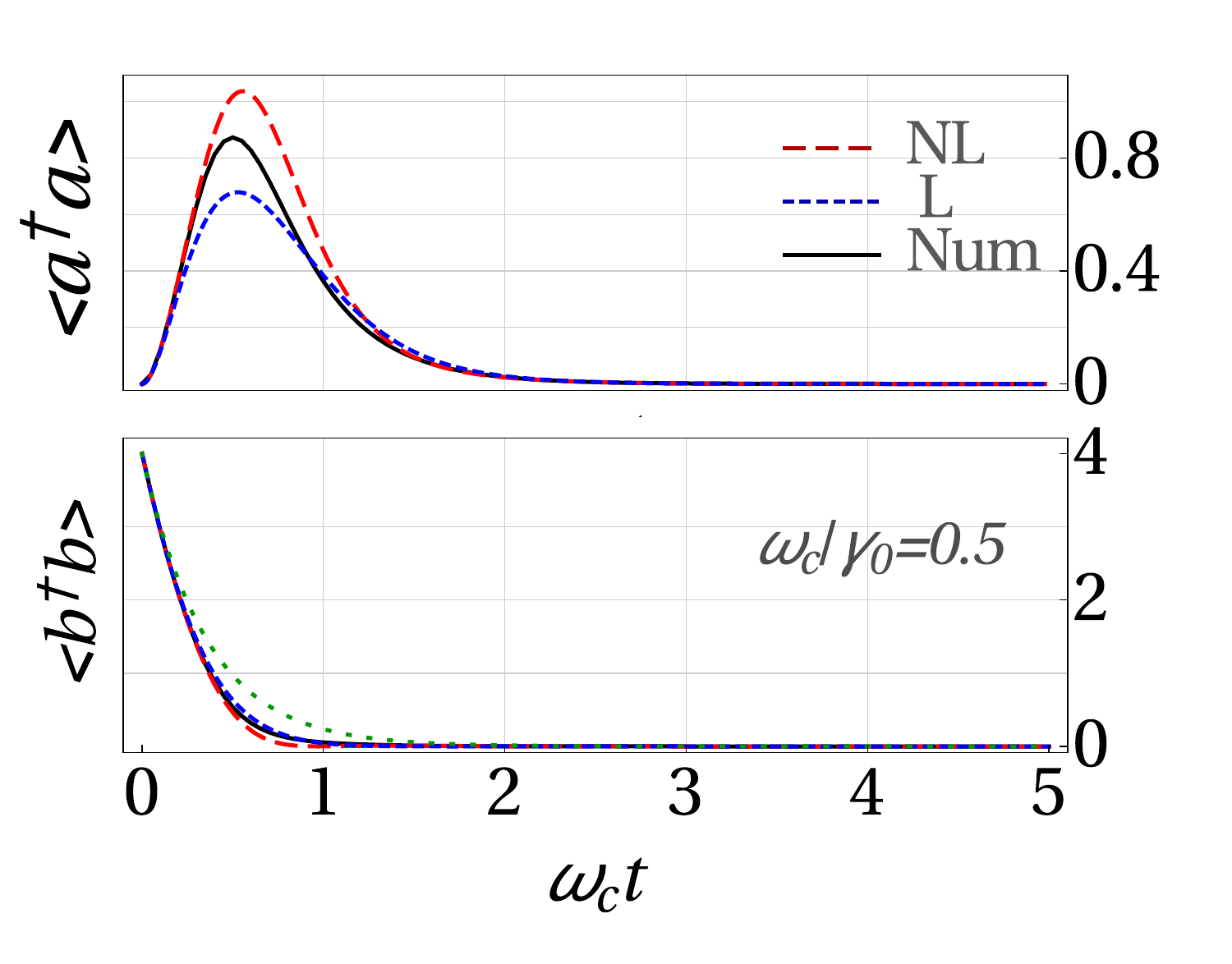}}\\
\subfigure[\label{fig:FreeB2G0p2}]
{\includegraphics[width=0.45\linewidth]{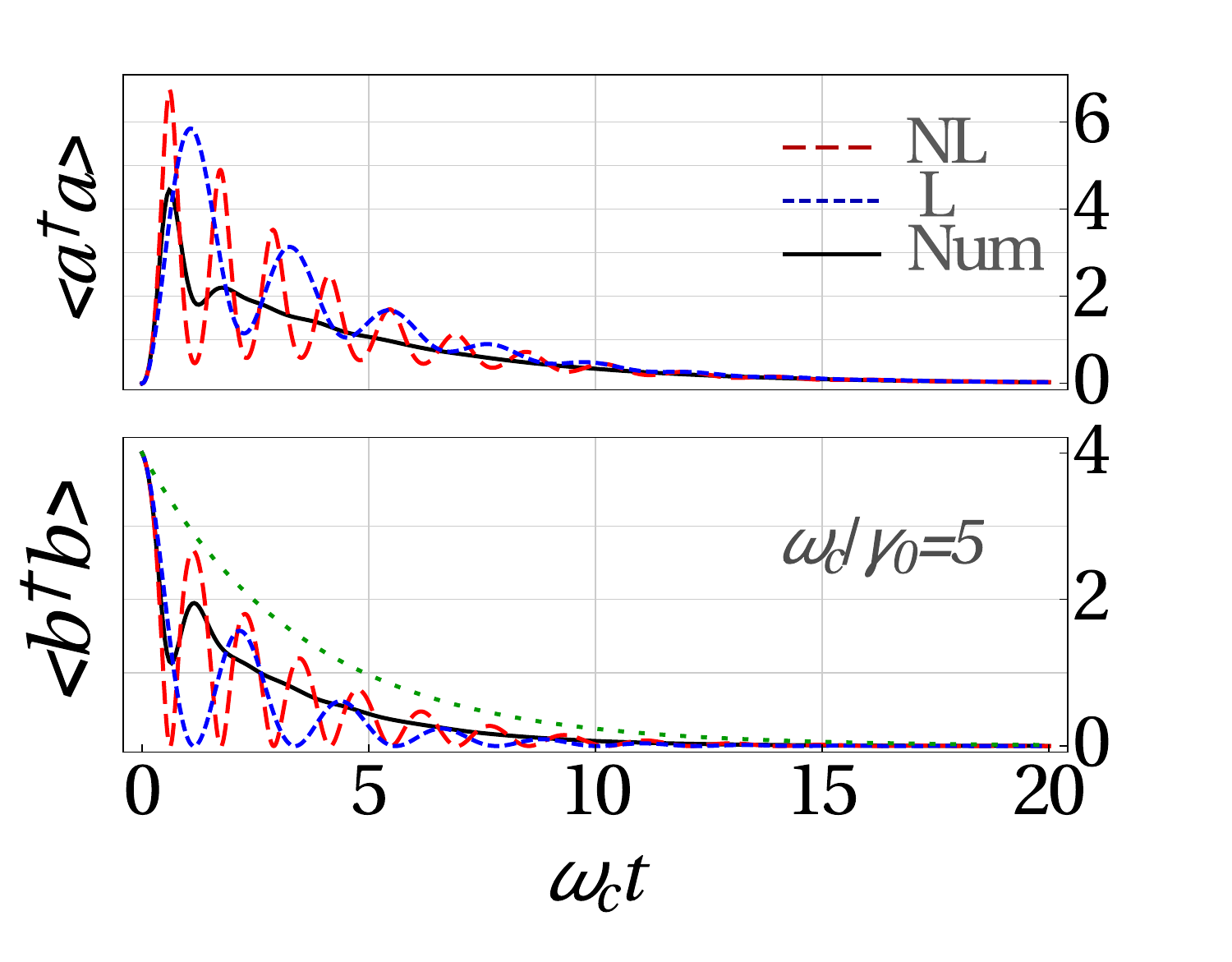}}\qquad
\subfigure [\label{fig:FreeB2G0p1}]
{\includegraphics[width=0.45\linewidth]{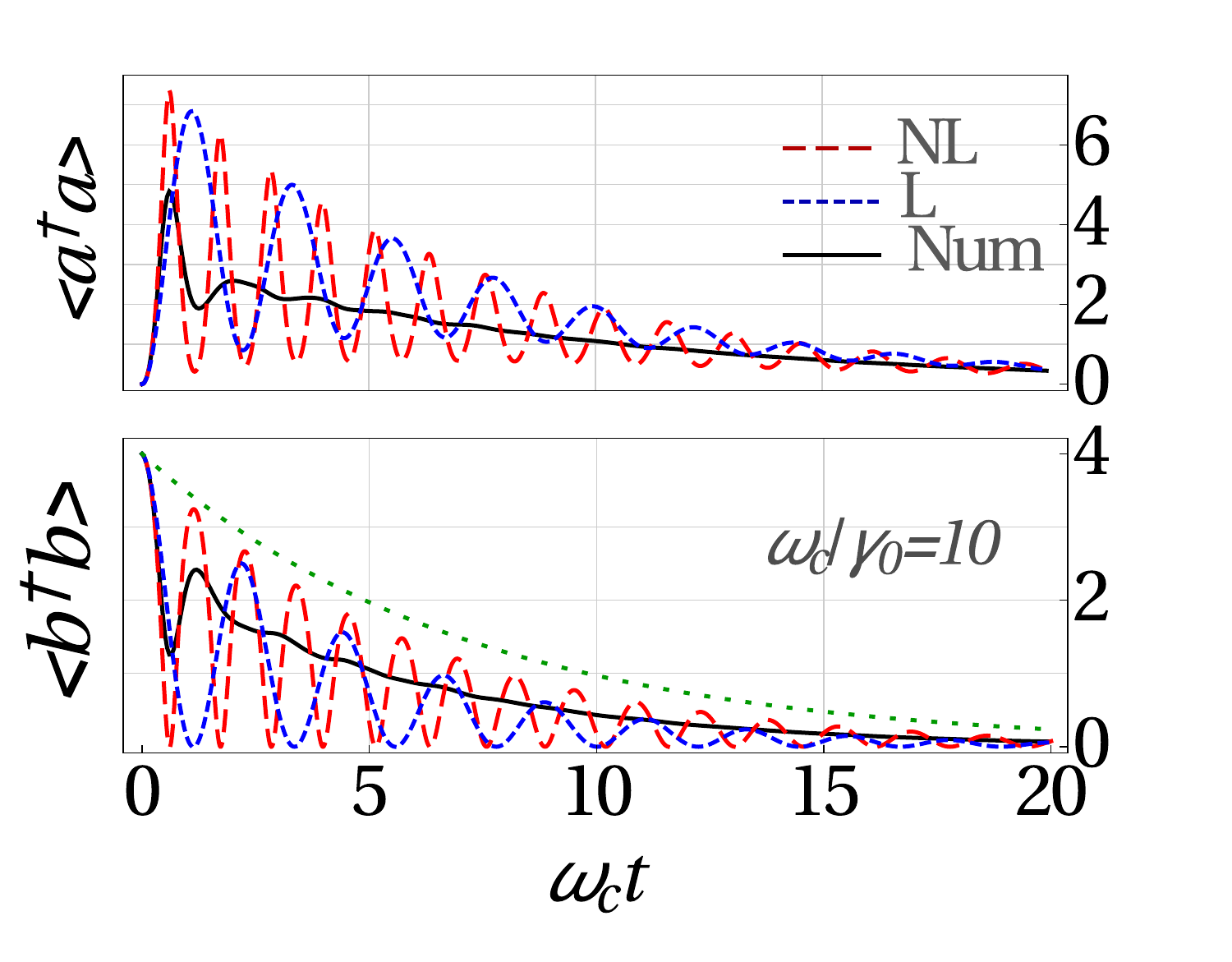}}
\caption{Free evolution of the interacting cavity field-mirror system. At the initial time $t=0$, the cavity field is in its vacuum state, while the mechanical oscillator is in a coherent state of amplitude $b_0=2$. Solutions are given in terms of the number of photons in the cavity $\langle \hat{a}^\dagger \hat{a}\rangle$ and of mechanical oscillator quanta $\langle \hat{b}^\dagger \hat{b}\rangle$. The different panels (a-d) are for growing values of $\omega_c/\gamma_0=0.1, 0.5, 5, 10$. In each panel, the solution (L) of the linearized equations is plotted as a blue fine dashed line, the solution (NL) of the nonlinear mean-field equations is shown as a dashed red line, and the full numerical solution (ME) of the master equation Eq.~\eqref{MasterEq} is shown as a continuous (black) line. In the panels for $\langle \hat{b}^\dagger \hat{b}\rangle$, the dotted green lines (not distinguishable in panel (a)) show the evolution of the mirror in the absence of opto-mechanical coupling to the cavity field, $\omega_c=0$.}
\label{fig:FreeEvol}
\end{figure*}

We start in this section the study of back-reaction effects, considering first the case of the free evolution of the mirror: the physical idea is that the cavity is prepared in its vacuum state, while the mirror is prepared in a coherent state with a given amplitude. Starting from this state, the system is then let evolve in the absence of any external drive $F=0$. The dissipative nature of the evolution will eventually bring it back to the ground state with all fields being in their vacuum state, but the intermediate dynamics will carry interesting signatures of the dynamical Casimir and of the back-reaction effects. 

To investigate this physics, we go back to the full set in Eqs.~(\ref{EqMotForc-1}-\ref{EqMotForc-3}). In absence of the external drive, that is for $F_0=0$, there is no advantage in moving to the rotating frame for the operators. We start from the simplest and most relevant regime of a small initial perturbation from the ground state, in which case the equations can be linearized into the form
\begin{align}
	\frac{db}{dt}&=-i\omega_b b -\frac{\gamma_b}{2}b-i\omega_c q,
\label{EqMotLin-1}\\
		\frac{dq}{dt}&=-\left(\gamma_a-2i\omega_a\right) q-2i\omega_c b,
\label{EqMotLin-2}
\end{align}
which describe a damped oscillating evolution for both the mirror and the field amplitude starting from the initial conditions $b(0)=b_0$ and $q(0)=0$. These equations of motion can be analytically solved and, in the resonant case $\omega_b=2\omega_a$, provide the solutions
\begin{align}
	b(t)&=e^{-i\omega_b t-\gamma_1 t/2}\frac{b_0}{2\omega_d}\left(i\left(\gamma_a-\frac{\gamma_b}{2}\right)\sin(\omega_d t)+2\omega_d\cos(\omega_d t)\right),\label{b-lin-sol}\\
	q(t)&=e^{-2i\omega_a t -\gamma_1 t/2}\frac{b_0}{4\omega_d\omega_c}\left(\left(\gamma_a-\frac{\gamma_b}{2}\right)^2-4\omega_d^2\right)\sin(\omega_d t). \label{q-lin-sol}
\end{align}
which show a complex temporal envelope modulating the free oscillations of $b(t)$ and $q(t)$ at $\omega_b=2\omega_a$. A similar expression can be obtained also for the average number of photons $n_a$, but we do not report here because it is quite involved and not that instructive. For the sake of compactness, we have used the shorthands $\gamma_1=\gamma_a+\gamma_b/2$ for the averaged dissipation rate and 
\begin{equation}
\omega_d=\sqrt{2\omega_c^2-\left(\gamma_a-\gamma_b/2\right)^2/4}\,,
\end{equation}
where the positive solution of the square root is here meant to be taken.
As a key result of this work, and in agreement with the conclusions of the recent work~\cite{Savasta-PRX-2018}, we easily see that two regimes can be identified, depending on the relative values of the interaction frequency $\omega_c$ and the dissipation rates $\gamma_a$ and $\gamma_b$, i.e. the real vs. imaginary nature of $\omega_d$. 

For $\omega_c>\left(\gamma_a-\gamma_b/2\right)/(2\sqrt{2})$, $\omega_d$ is real and energy is periodically exchanged between the mirror and the optical mode of the cavity, before being eventually damped with an exponential law on a longer timescale. In the opposite case, damping is so large that $\omega_d$ is purely imaginary and the amplitude of the mirror oscillations is monotonically damped out. Of course, the resulting damping rate gets contributions from the bare decay rates $\gamma_{a,b}$ as well as from the back-reaction effect. Since the dynamical Casimir emission is suppressed for substantial values of the mirror-cavity detuning $|\omega_b-2\omega_a| \gg \gamma_{a,b}$, the back-reaction contribution can be isolated by comparing the values of the damping rate that are observed in the two cases when the cavity is tuned respectively on- or far-off resonance from the mirror.

This linearized approach holds for weak initial amplitudes $b_0^2\ll \gamma_{a}^2/\omega_c^2$, so that the nonlinear terms in the motion equations are negligible. In more general case, the full quantum nonlinear equations (\ref{EqMotForc-1}-\ref{EqMotForc-3}) should be considered. For small values of $\omega_c/\gamma_{a,b}$, one can expect that nonlinear mean-field equations (\ref{EqMotSoft-1}-\ref{EqMotSoft-3}) should provide a reasonable approximate description.

These analytical expectations are validated in Fig.~\ref{fig:FreeEvol} that shows the free evolution of the system starting from $b_0=2$ and the cavity field in its vacuum state. The panels (a-d) refer to growing values of coupling strength, $\omega_c/\gamma_0=0.1,\,0.5,\,5,\,10$. In each panel, the different curves show the full numerical solution of the master equation (black solid line), the solution of the linearized equations (blue dotted) and the solution of the nonlinear mean-field equations (red dashed). The dotted green lines in the panels for $\langle \hat{b}^\dagger \hat{b}\rangle$ show the bare damping of the mechanical oscillator at $\gamma_b$. For simplicity we have assumed equal dissipation rates for both the cavity and the mechanical oscillator $\gamma_a=\gamma_b=\gamma$, so that $\gamma_0=\gamma/\sqrt{2}$ and $\gamma_1=3\gamma/2$. With this choice, one has $\omega_d=\left(2\omega_c^2-\gamma^2/16\right)^{1/2}$, so that the condition separating the over- and under-damped regime is $\omega_c/\gamma_0=1/4$.  

In panels (a-b), we illustrate the over-damped regime of weak opto-mechanical coupling: while the mechanical oscillator performes a monotonic decay towards its ground state, the cavity field is initially excited by the dynamical Casimir effect, then the photons are lost by dissipation. Consequences of back-reaction are anyway visible in the decay rate of the mirror oscillation, that is reinforced compared to its bare value $\gamma_b$ (green dotted line). The quantitative importance of this effect grows with $\omega_c^2$, which makes it clearly visible in Fig.~\ref{fig:FreeEvol}(b), but almost invisible in panel (a). In both these panels, the agreement of the different approximations to the full numerical solution is very good and the discrepancy gets smaller as $\omega_c/\gamma_0$ is decreased.

In panels (c-d), we illustrate the under-damped regime where a continuous and periodic transfer of energy occurs between the mirror and the field and viceversa. The time scale on which such a conversion takes place can be estimated from the analytical theory to be on the order of $1/\omega_d$. Because of the losses, the system then decays towards the vacuum state on a time scale set the characteristic time $1/\gamma_1$. As expected, the agreement between the analytical approximations and the full numerics is worse for larger $\omega_c$. Given the relatively large initial value of $b$ chosen here, the linearized approach provides inaccurate results. The nonlinear mean-field equations are however able to reasonably capture the oscillation frequency. Quantum fluctuations and correlations are then responsible for the quick damping of the oscillations shown in the full numerics.

\subsection{Driven-dissipative steady-state under a monochromatic drive\label{subsec:Driven}}

After having discussed the free evolutiom of the system under the combined effect of the losses and the dynamical Casimir emission, we now turn to the driven-dissipative dynamics when the system is continuously driven by a monochromatic drive acting on the mirror. As we have done in the previous section, the full numerical results will be compared to the mean-field nonlinear equations (\ref{EqMotSoft-1}-\ref{EqMotSoft-3}): as compared to the pure mean-field theory based on one-operator expectation values of Sec.\ref{subsec:ParOsc}, these equations explicitly include the relevant two-operator quantities that enter into the DCE, in particular $q=\langle \hat{a}^2\rangle$.

\subsubsection{Linear regime\label{subsubsec:DrivenLinear}}

We start from the case where the strength of the external drive is small enough that the system is slightly perturbed from the vacuum. In this regime, an analytical solution for the response of the mirror can be obtained by linearizing the equations of motion, which gives
\begin{equation}
	b(\omega)=\mathcal{R}(\omega)F_0,
\label{b-SteadyAmpl}
\end{equation}
where $\mathcal{R}(\omega)$ is the response function of the oscillator 
\begin{equation}
	\mathcal{R}(\omega)=-\frac{\left(\Delta_q+i\gamma_a\right)}{\left[\Delta_b\Delta_q+i\left(\gamma_a\Delta_b+\frac{\gamma_b}{2}\Delta_q\right)-\left(2\omega_c^2+\gamma_0^2\right)\right]}\,.
\label{b-SteadyAmpl-general}
\end{equation}
This formula is one of the key results of our work. In the completely resonant case $\omega_b=2\omega_a$, it simplifies as
\begin{equation}
	\mathcal{R}(\omega)=-\frac{\left(\Delta+i\gamma_a\right)}{\left[\Delta^2+i\gamma_1\Delta-\left(2\omega_c^2+\gamma_0^2\right)\right]}\,.
\label{ResponseFunc}
\end{equation}
where $\Delta=\omega-\omega_b=\omega-2\omega_a$, $\gamma_0=\sqrt{\gamma_a\gamma_b/2}$ and $\gamma_1=\gamma_a+\gamma_b/2$.
As expected, in the limit $\omega_c\rightarrow 0$ of a vanishing opto-mechanical interaction, the response function reduces to the response \eqref{BareRespFunc} of the bare oscillator. 

For small $\omega_c\ll \gamma_1$, and assuming for simplicity $\gamma_b \ll \gamma_a$, 
the response \eqref{ResponseFunc} takes the Lorentzian form
\begin{equation}
	\mathcal{R}(\omega)=-\frac{1}{\left[\Delta+i\left(\gamma_0^2+2\omega_c^2\right)/\gamma_1\right]},
\label{b-SteadyAmpl-limit}
\end{equation}
and the steady-state oscillation of the mirror has a (squared) amplitude
\begin{equation}
	\left |b(\omega)\right|^2=\frac{\left|F_0\right|^2}{\left[\Delta^2+\left(\frac{\gamma_0^2+2\omega_c^2}{\gamma_1}\right)^2\right]}\,.
\label{b2-SteadyAmpl-limit}
\end{equation}
Both these last formulas clearly show the back-reaction effect of the dynamical Casimir emission as a reinforced broadening of the mirror response,
\begin{equation}
 \Gamma\equiv \frac{\gamma_0^2+2\omega_c^2}{\gamma_1}\simeq \gamma_b+\frac{2\omega_c^2}{\gamma_a}~:
\end{equation}
while the first term is the bare damping of the mechanical oscillator, the second term accounts for the damping due to the creation of photon pairs out of the vacuum. Since the dynamical Casimir effect is dramatically suppressed far away from resonance $|\omega_b-2\omega_a|\gg \gamma_{0,1}$, the back-reaction contribution can be extracted just by looking at the dependence of the linewidth on the cavity-mirror detuning $\omega_b-2\omega_a=\Delta_q-\Delta_b$.

For arbitrary values of $\omega_c$, the squared amplitude of the mirror oscillations follows directly from Eq.~\eqref{ResponseFunc},
\begin{equation}
	\left |b(\omega)\right|^2=\frac{\left(\Delta^2+\gamma_a^2\right)}{\left(\Delta^2-(2\omega_c^2+\gamma_0^2)\right)^2+\gamma_1^2\Delta^2} \left|F_0\right|^2\,.
\label{b2-SteadyAmpl}
\end{equation}
For strong values of the coupling $\omega_c\gg \gamma_{a,b}$, the periodic energy exchange between mechanical and optical modes predicted in the previous section manifests itself in a complex response spectrum showing a pair of Lorentzian peaks of width $\gamma_1$ separated by a splitting approximately given by $2\sqrt{2} \omega_c$,
\begin{equation}
	\left |b(\omega)\right|^2=\frac{2\omega_c^2}{\left(\Delta^2-2\omega_c^2\right)^2+2\omega_c^2\gamma_1^2} \left|F_0\right|^2\,.
\label{b2-SteadyAmpl2}
\end{equation}

\subsubsection{Nonlinear mean-field regime\label{subsubsec:DrivenNonLinear}}

For higher strength of the drive, the response of the system need to be computed at the classical level by taking into account the nonlinear character of the system, encoded in the Eqs.~\eqref{EqMotSoft-1}-\eqref{EqMotSoft-3}. Since we are interested in the stationary state of the system, we pose the time derivatives to zero here. To analytically tackle the nonlinear equation, we combine the first and the second ones to find the steady-state for the mirror oscillation amplitude and for the field fluctuations as a function of the average number of photons in the cavity $n_a$,
\begin{align}
	b&=\frac{\left(\Delta+i\gamma_a\right) F_0}{\left[2\omega_c^2\left(1+2n_a\right)-\left(\Delta^2-\gamma_0^2\right)\right]-i\Delta\gamma_1},
\label{bNonLin-SteadyAmpl}\\
	q&=\frac{2\omega_c\left(1+2n_a\right)F_0}{\left[2\omega_c^2\left(1+2n_a\right)-\left(\Delta^2-\gamma_0^2\right)\right]-i\Delta\gamma_1}.\label{qNonLin-SteadyAmpl}
\end{align}

\begin{figure}[htbp]
\centering%
\includegraphics[width=0.9\linewidth]{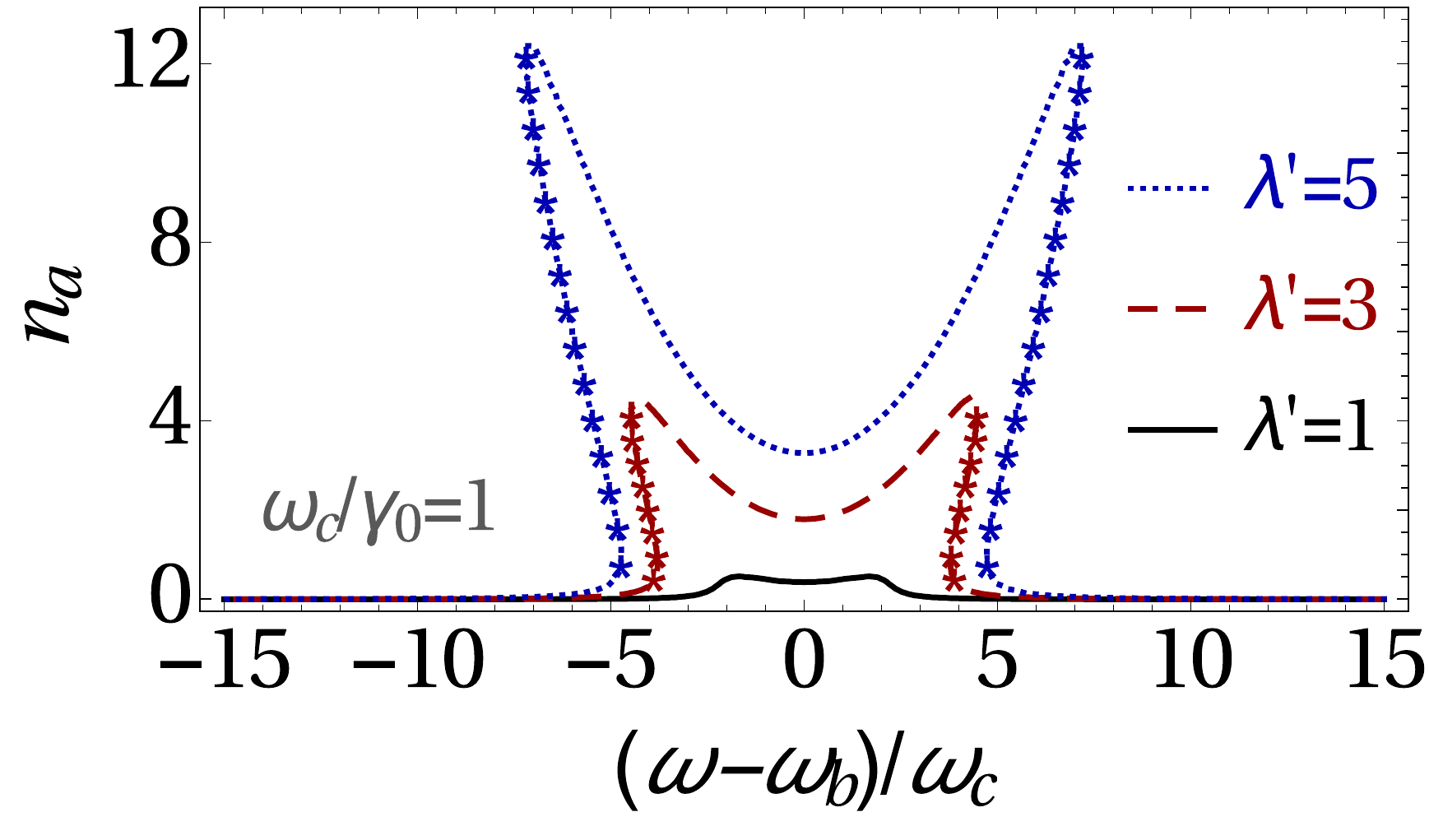}
\caption{Stationary state in the presence of a monochromatic mechanical drive acting on the mirror. Number $n_a$ of photons in the cavity as a function of the drive frequency $\omega$. All curves are for the resonant case $\omega_b=2\omega_a$, while different values $\lambda'=F_0/(F_0^{\text{th}})'=1,\,3,\,5$ for the drive amplitude are used. The presence of multiple solutions for the same drive frequency and amplitude indicates multistable behaviours, the dynamically unstable branches being highlighted by symbols.}
\label{fig:Bistability}
\end{figure}

Here we posed $\omega_b=2\omega_a$ (that is $\Delta_a=\Delta_b=\Delta$). From the third equation we then find a condition on $x\equiv\left(1+2n_a\right)$ in the form of a third order polynomial equation,
\begin{multline}
	\omega_c^4 x^3-\omega_c^2\left[\omega_c^2+\left(\Delta^2-\gamma_0^2\right)\right]x^2+\left[\frac{\left(\Delta^2-\gamma_0^2\right)^2}{4}+\omega_c^2\left(\Delta^2-\gamma_0^2\right)\right.\\
	\left.+\frac{\Delta^2}{4}\gamma_1^2-4 F_0^4\right]x-\frac{1}{4}\left(\left(\Delta^2-\gamma_0^2\right)^2+\Delta^2\gamma^2\right)=0.
\label{xNonLin-SteadyAmpl}
\end{multline}
which can be easily solved by numerical means. The solution then provides the amplitude of the mirror and field oscillations through Eqs.~\eqref{bNonLin-SteadyAmpl} and \eqref{qNonLin-SteadyAmpl}. 

Because of the nonlinear nature of the problem, multiple (up to three) solutions could exist for these equations, depending on the values of the strength and frequency of the drive. Such \emph{multistability} effects are well-known in optics and a simplest example in our context is illustrated in Fig.~\ref{fig:Bistability}: depending on the drive frequency, one or two stable solutions can be found, as well as dynamically unstable ones. The splitting of the two tilted peaks is due to a nonlinear Rabi coupling between the mechanical and optical degrees of freedom~\cite{IC-GCLR-PRB1998}, and, for $n_a\gg 1$ can be estimated from \eqref{xNonLin-SteadyAmpl} to be approximately $2\omega_c\,n_a^{1/2}$.

\paragraph{Modified parametric oscillator\label{subsubsec:DrivenModParOsc}}
Hu-MOF-2013
This general theory can be successfully used to study the dynamical Casimir emission and the back-reaction effect in the nonlinear regime. 
For simplicity we restrict to the fully resonant case $\omega=\omega_b=2\omega_a$ and we give a special attention to the field fluctuations, taken into account in our theory at the level of the averages of the $\hat{a}^2$ operator. Setting $\Delta=0$, Eqs.~(\ref{bNonLin-SteadyAmpl}-\ref{qNonLin-SteadyAmpl}) simplify to
\begin{align}
	b&=\frac{i\gamma_a \omega_c F_0}{2\omega_c^2 x+\gamma_0^2},
\label{bNonLin-SteadyAmpl-Om0}\\
	q&=\frac{2\omega_c^2 F_0}{2\omega_c^2 x+\gamma_0^2},\label{qNonLin-SteadyAmpl-Om0}
\end{align}
while \eqref{xNonLin-SteadyAmpl} reduces to
\begin{multline}
	\omega_c^4x^3-\omega_c^2\left(\omega_c^2-\gamma_0^2\right)x^2\\\left(\frac{\gamma_0^4}{4}-\omega_c^2\gamma_0^2-4\omega_c^4 F_0^2\right)x-\frac{\gamma_0^4}{4}=0.
\label{xNonLin-SteadyAmpl-Om0}
\end{multline}
In the $F_0\to 0$ limit, this set of equation admits the explicit solutions
\begin{equation}
	b=\frac{i \omega_c\gamma_a F_0}{2\omega_c^2+\gamma_0^2},\quad q=\frac{2\omega_c^2 F_0}{2\omega_c^2+\gamma_0^2},\quad n_a=\frac{8\omega_c^4 F_0^2}{\left(2\omega_c^2+\gamma_0^2\right)^2}
\label{F0to0}
\end{equation}
that fully recover the result of the linearized equations \eqref{EqMotSoft-1}-\eqref{EqMotSoft-3}. This is immediately seen, for example, by comparing the expression for $b$ in Eq.~\eqref{F0to0} with the one in \eqref{b-SteadyAmpl} and \eqref{ResponseFunc} for $\Delta=0$.  

\begin{figure}[htbp]
\centering%
\includegraphics[width=0.9\linewidth]{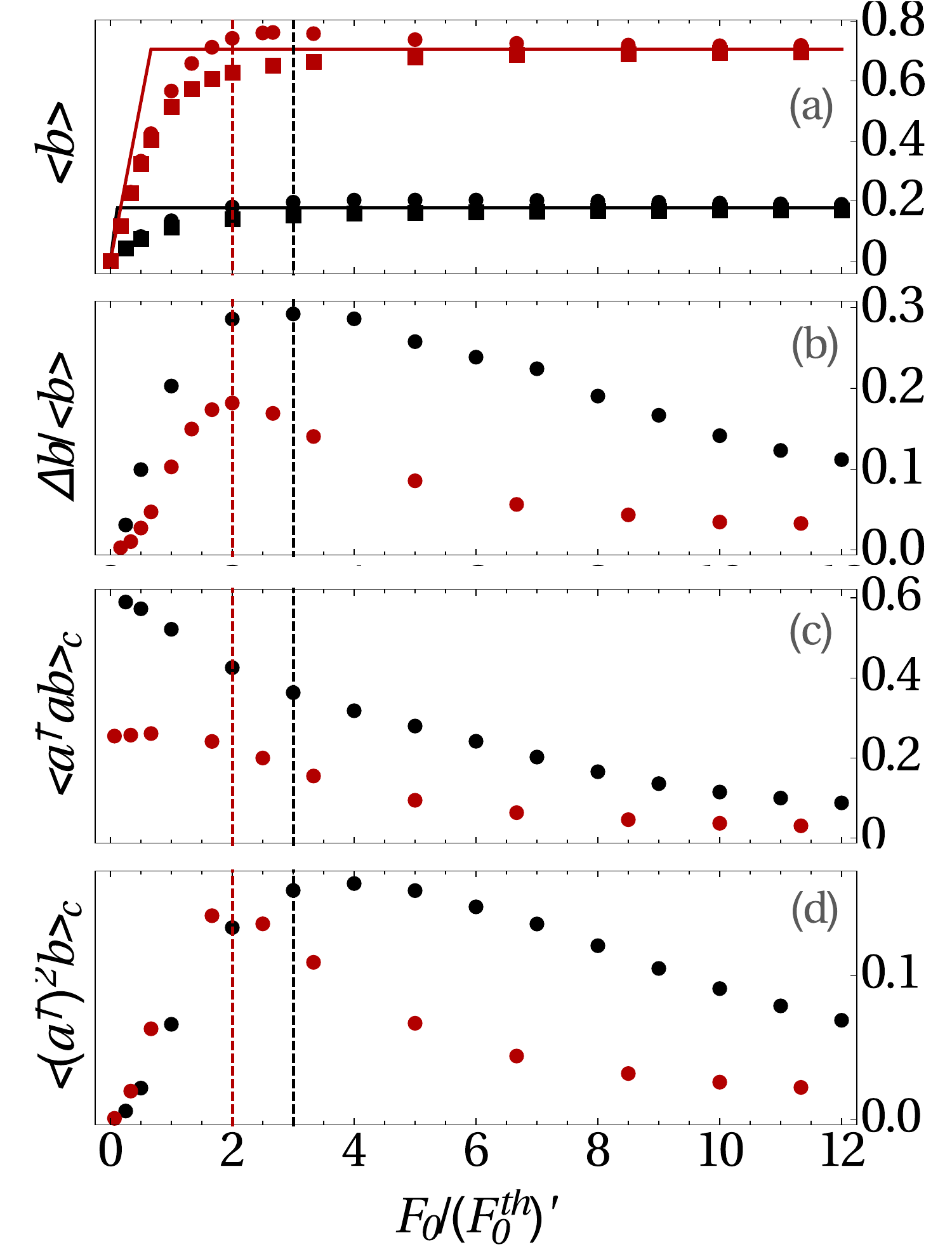}
\caption{Stationary state of the driven-dissipative evolution in the presence of a monochromatic drive acting on the mirror in a fully resonant condition $\omega=\omega_b=2\omega_a$. Panel (a) shows the amplitude of the mirror oscillations as a function of the drive amplitude. The solid line is the mean-field solution of (\ref{MeanFieldA}-\ref{MeanFieldB}), squares show the analytical nonlinear expression \eqref{xNonLin-SteadyAmpl-Om0} and the circles indicate the full numerical prediction of the master equation. Different colors refer to different values of the optomechanical coupling, $\omega_c/\gamma_0=0.5$ (red) and $2$ (black). Panel (b) shows the relative deviation between the analytical nonlinear solutions and the full numerical solution. Panels (c,d) show the numerical solutions for the normalized correlations between the field and the mirror, as defined in Eqs.~\eqref{Correlations}. Vertical dashed lines indicate the points of maximum deviation $\Delta b/\left<b\right>$.}
\label{fig:FigB}
\end{figure}

In the opposite limit $F_0\to\infty$, the nonlinear mean-field equations predict for the stationary state of the system
\begin{equation}
	b=\frac{i\gamma_a}{4\omega_c},\quad q=F_0,\quad n_a=F_0~:
\label{F0toInf}
\end{equation}
the equal expressions for $q$ and $n_a$ suggest that in this regime the cavity field is in a coherent state and its amplitude recovers the mean-field prediction \eqref{BAboveTr}. 

The different dependence on the strength $F_0$ of the drive appearing in Eqs.~\eqref{F0to0} and \eqref{F0toInf} is a hint of the parametric oscillator transition. In order to estimate the amplitude of the drive at which the crossover between the below- and above-threshold regimes takes place, we equate the amplitude of the mechanical oscillations as predicted in Eqs.~\eqref{F0to0} and \eqref{F0toInf}, obtaining the threshold value $\left(F{_{0}^{\text{th}}}\right)'=(1/2)\left(\omega_c+\gamma_0^2/2\omega_c\right)$. In the limit $\omega_c\to 0$, this expression reduces to the critical value $F_0^{\text{th}}$ predicted by the mean-field theory. Such a transition is illustrated in Fig.~\ref{fig:FigB}(a), where is shown the solution for $b$ as a function of $F_0$, for the values $\gamma_a/\sqrt{2}=\gamma_b/\sqrt{2}=\gamma/\sqrt{2}=0.5,\,2$ (that is for $\gamma_0=0.5,\,2$ respectively). A close analysis of the figure reveals the existence of three distinct regimes depending on the value of the ratio $\lambda'\equiv F_0/\left(F{_{0}^{\text{th}}}'\right)$ of the drive strength. In contrast to the pure mean-field theory discussed in Sec.\ref{subsec:ParOsc}, the transition between the different regimes is not sharp but is smoothened out by quantum fluctuations.

The three regimes correspond to i) \emph{below} $(\lambda'\ll 1)$, ii) \emph{above} $(\lambda'\gg 1)$ and iii) \emph{around} $(\lambda'\simeq 1)$ threshold. The solutions in Eq.~\eqref{F0to0} refers to the first of these regions (regime (i)). The most interesting feature is that the quantum fluctuations due to the mirror-field interaction decrease the slope of $b$ as a function of $F_0$ with respect to the mean-field prediction in Eq.~\eqref{BelowTr} and this deviation grows with $\omega_c$. All these elements confirm the origin of this feature in the DCE emission of photons that increases the effective damping of the mirror via the back-reaction effect. Note also that, in this regime, the theoretical solution agrees very well with the (fully quantum) numerical one. This happens because, despite the quantum fluctuations are not negligible in this limit and the factorization of the correlations is not justified, the system is only weakly displaced from its vacuum state, and the correlations account for higher order terms in the infinitesimal displacement of the system above its vacuum state.

\begin{figure*}[htbp]
\centering%
\subfigure [\label{fig:spectrum1}]
{\includegraphics[width=0.45\linewidth]{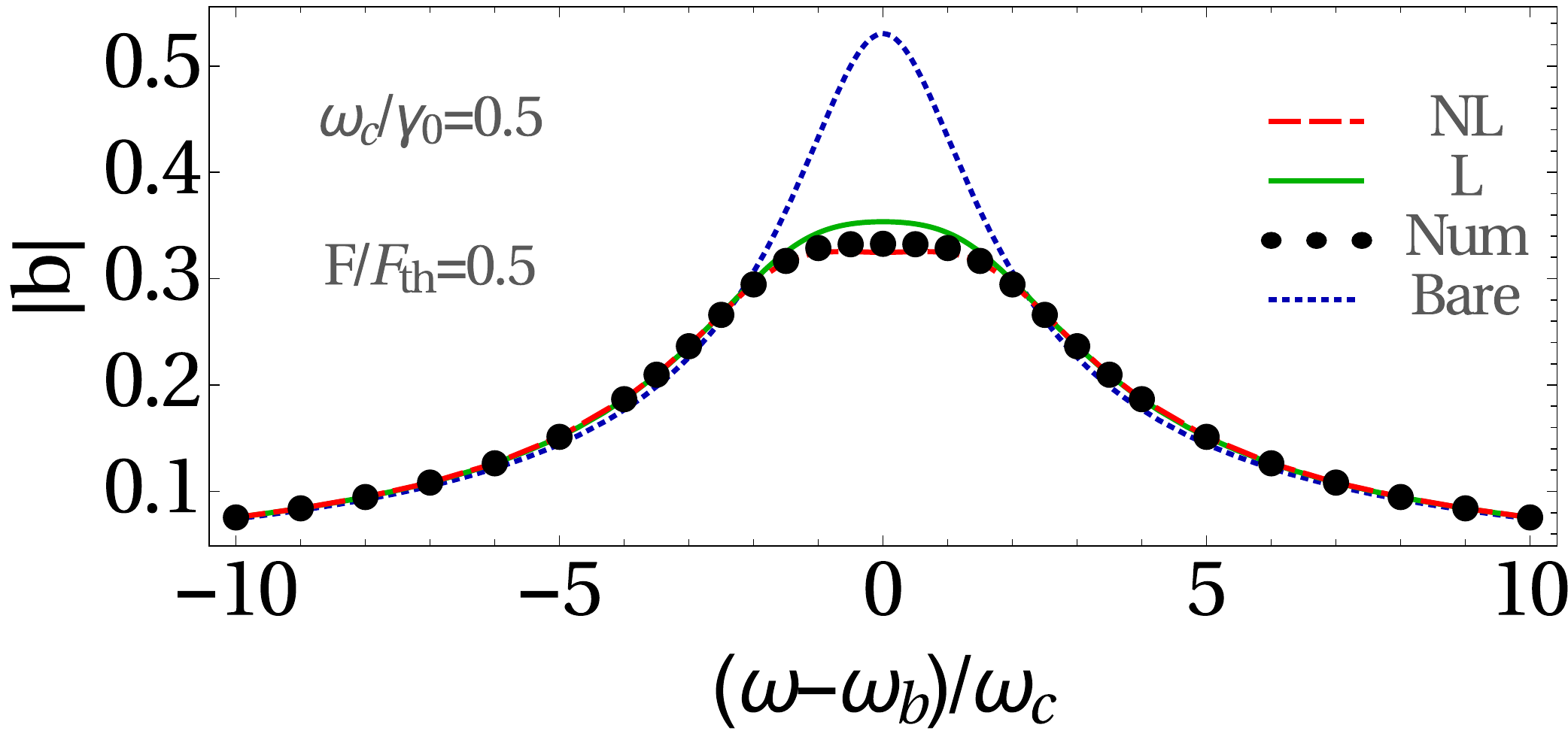}}\quad
\subfigure [\label{fig:spectrum2}]
{\includegraphics[width=0.45\linewidth]{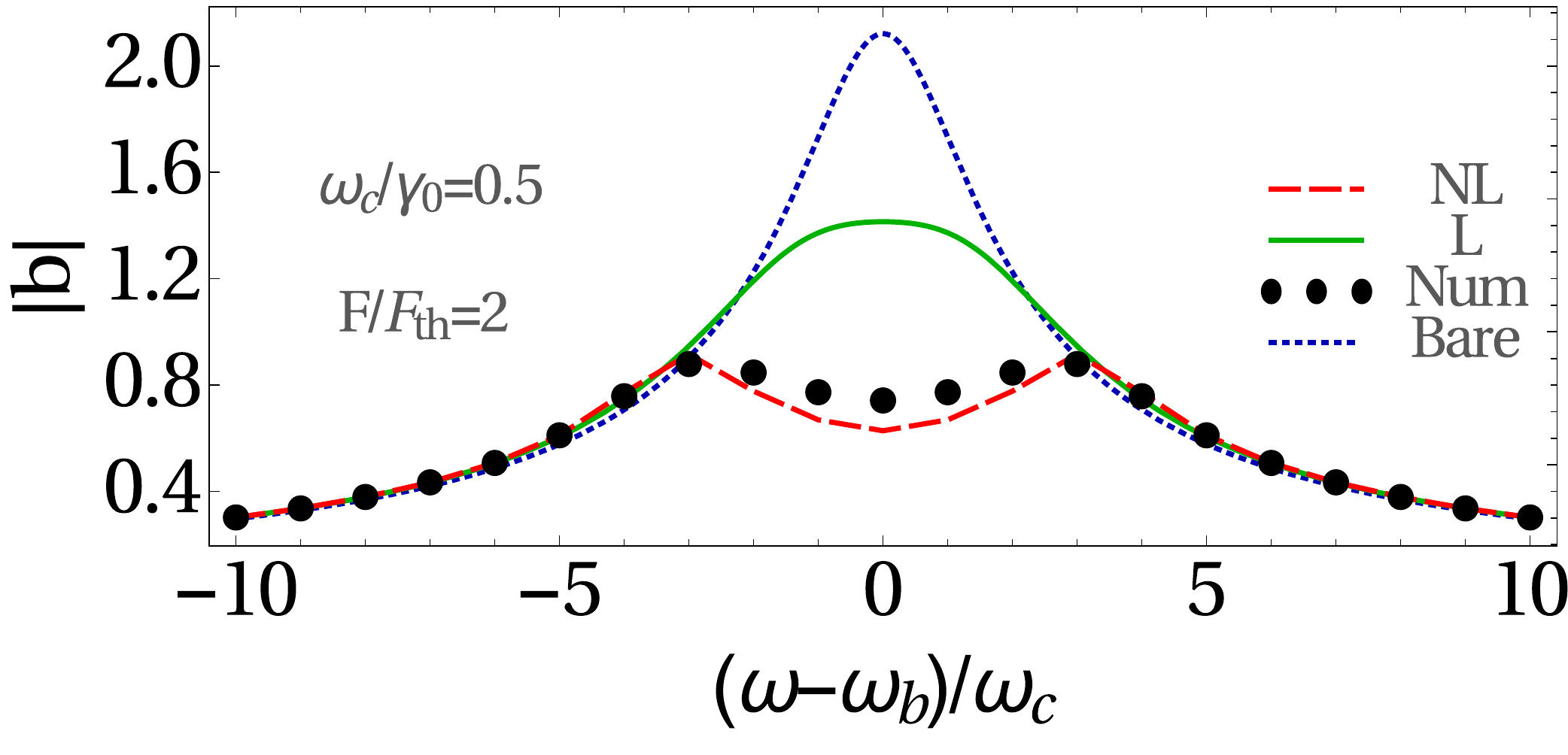}}\\
\subfigure [\label{fig:spectrum3}]
{\includegraphics[width=0.45\linewidth]{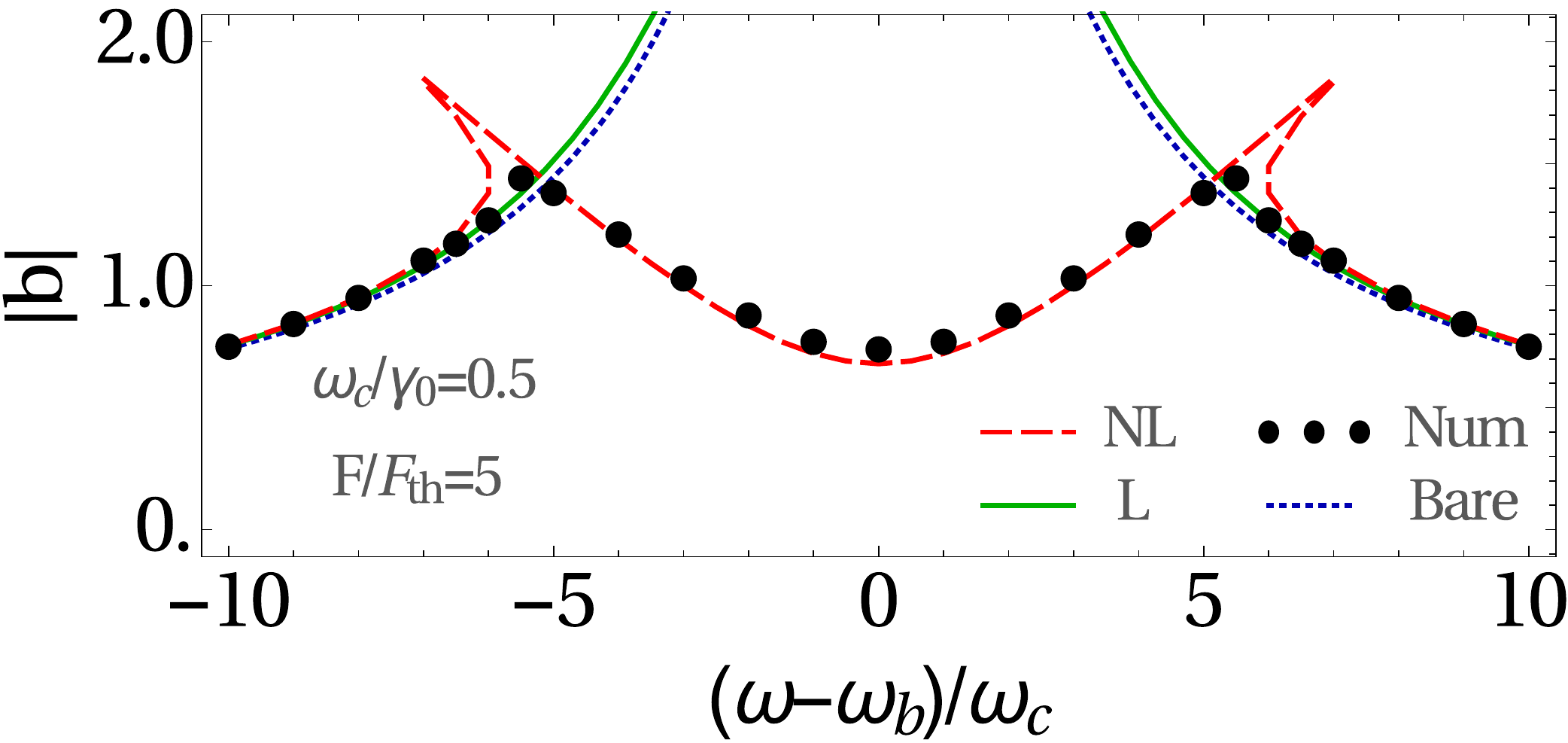}}\quad
\subfigure [\label{fig:spectrum4}]
{\includegraphics[width=0.45\linewidth]{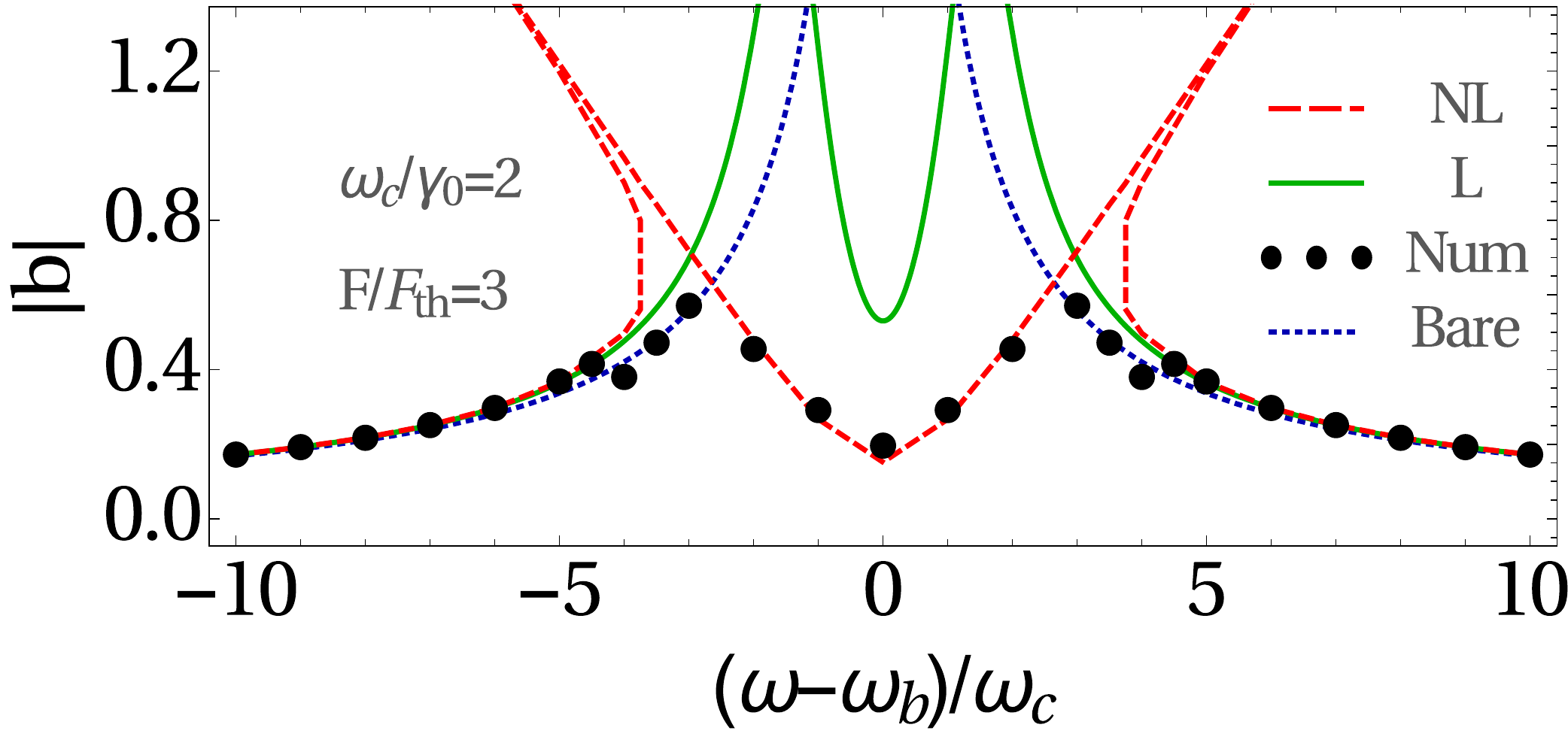}}
\caption{Steady-state amplitude of the mirror oscillation as a function of the frequency $\omega$ of the monochromatic drive.  We consider the mechanical oscillator resonant with the cavity field: $\omega_b=2\omega_a$, and use different values for the optomechanical coupling and of the drive amplitude. Panels (a-c): $\omega_c/\gamma_0=0.5$ and $F_0/\left(F_0^{\text{th}}\right)'=0.5$ (a), $F_0/\left(F_0^{\text{th}}\right)'=2$ (b), $F_0/\left(F_0^{\text{th}}\right)'=5$ (c). Panel (d): $\omega_c/\gamma_0=2$ and $F_0/\left(F_0^{\text{th}}\right)'=3$. The black symbols indicate the numerical solution of the master equation \eqref{MasterEq}. The prediction of the linearized theory is shown as a solid green line. The prediction of the nonlinear mean-field model is shown as a dashed red line. The response of the bare mirror for a vanishing optomechanical coupling is shown as a dotted blue line.}
\label{fig:spectrum}
\end{figure*}

Above the parametric oscillator transition (regime (iii)), the coherent oscillations of the mirror generated by the driving force are so large to induce a self-supported coherent oscillation in the cavity field as well. In the DCE context, such oscillations were observed in \cite{Wilson-PRL-2010} and must, of course, be distinguished from the quantum-fluctuation-induced excitation that is observed in the cavity in the regime (i) below the transition~\cite{Wilson-DCE-Analog-2011}. Also in this regime (iii), the mean-field solution agrees well with both the theoretical and the numerical solutions: the system behaves in fact classically and quantum effects have a negligible impact on the dynamics. 

In the region (ii) in between these two limits, that is for values of the order $F_0\sim \left(F{_{0}^{\text{th}}}\right)'$, the quantum fluctuations have non-negligible effects on the properties of the system, whose dynamics significantly deviates from the prediction of the theoretical model developed in the previous sections. These considerations are supported fom the numerical results in Fig.~\ref{fig:FigB}(c-d), where the normalized correlations
\begin{align}
&\left\langle\hat{n}_a\hat{b}\right\rangle_c\equiv \frac{\left\langle\hat{n}_a\hat{b}\right\rangle}{n_a b}-1\\
&\left\langle\left(\hat{a}^\dag\right)^2\hat{b}\right\rangle_c\equiv \frac{\left\langle\left(\hat{a}^\dag\right)^2\hat{b}\right\rangle}{q^* b}-1
\label{Correlations}
\end{align}
are plotted as a function of the drive strength. From the same figures, we also confirm the expectation that the stronger is the optomechanical coupling compared to the loss rates $\omega_c/\gamma_0$, the stronger is the effect of the quantum correlations and thus the larger are the deviations of the analytical results from the fully quantum numerical solution (Fig.~\ref{fig:FigB}(b)).

\paragraph{Spectral response\label{subsubsec:DrivenSpectr}}

After having characterized the general features of the parametric transition in the fully resonant case, we now discuss the response of the mirror as a function of the drive frequency. In the linear regime of a weak drive, we obtained in Eqs.~\eqref{b2-SteadyAmpl} that the linewidth of the response function gets an additional contribution from the back-reaction effect of the DCE emission. 
The same conclusions can be drawn from the analysis of the more general nonlinear set of Eqs.~(\ref{EqMotSoft-1}-\ref{EqMotSoft-3}), despite in this case an explicit solution for the response function cannot be obtained. 

In Figs.~\ref{fig:spectrum}, these prediction are contrasted  with the corresponding numerical results. In panels (a-c) we consider the case of a relatively weak $\omega_c/\gamma_0=0.5$ and different values of the drive strength $F_0/\left(F_0^{\text{th}}\right)'=0.5,\,2,\,5$. We observe a good matching between the nonlinear analytical result and the numerical solution in the first and last cases, corresponding respectively to situations well below and well above the parametric oscillator transition. As expected, in the first case the linearized solution in Eq.~\eqref{b2-SteadyAmpl} also provides a good approximation to the response function. A sizeable deviation between the analytical and numerical results for the response function is instead observed in the intermediate case $F_0/\left(F_0^{\text{th}}\right)'=2$, that is the value of the drive strength for which the discrepancy between the analytical and numerical solutions for $\Delta=0$ was the largest in Fig.~\ref{fig:FigB}. 

The response for a larger value $\omega_c/\gamma_0=2$ of the optomechanical coupling is shown in Fig.~\ref{fig:spectrum4}. The drive amplitude $F_0/\left(F_0^{\text{th}}\right)'=3$ is again chosen to maximize the deviation between the analytical and numerical solutions for $\Delta=0$ shown in Fig.~\ref{fig:FigB}. As expected, by comparing Figs.~\ref{fig:spectrum2} and \ref{fig:spectrum4} we notice a better agreement between the two solutions in the case of a weaker opto-mechanical coupling.

\section{Circuit analogue\label{sec:Circuit}}
\begin{figure*}[t]
\centering%
\includegraphics[width=0.8\linewidth]{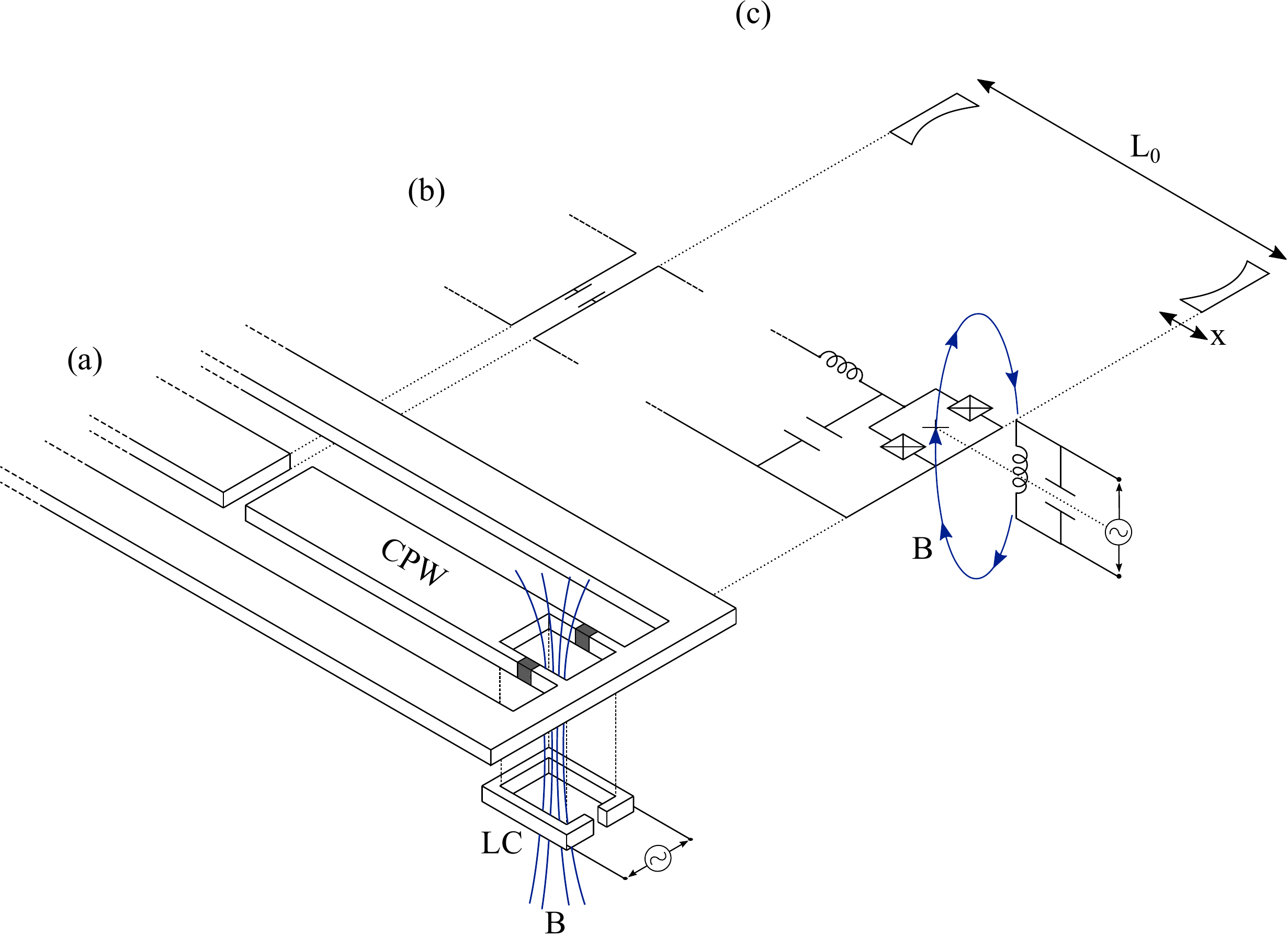}
\caption{a) Pictorial representation of $LC$ resonator magnetically coupled to the coplanar wave-guide. b) Sketch of the equivalent circuit. (c) Effective cavity with moving mirror in  correspondence of the SQUID.}
\label{fig:CircuitScheme}
\end{figure*}

As we have mentioned in the introduction, a direct observation of dynamical Casimir physics using a mechanically moving mirror is still facing great experimental difficulties. Even though the recent progresses in the miniaturization techniques nowadays allow to produce micro/nano-sized devices  in which the strength of the radiation pressure interaction is comparable to the other relevant energy scales of the system, an important challenge is still posed by the relatively low value of the characteristic mechanical frequencies, which hinders fulfillment of the resonance condition $\omega_b=2\omega_a$.

In the first observation of the dynamical Casimir effect~\cite{Wilson-DCE-Analog-2011}, this difficulty was circumvented by making use of an analog model based on a superconductor-based waveguide, where the role of the mirror is played by a SQUID device. In the analogy, the mechanical motion of the mirror in space is simulated by tuning the reflection phase of the SQUID via an externally imposed static magnetic field. 

In this section, we take inspiration from this experiment to propose a configuration where the mirror motion is not externally predetermined, but constitutes an independent degree of freedom of the system, dynamically coupled to the cavity field via the opto-mechanical Hamiltonian \eqref{Hint}. The basic idea is to replace the externally imposed magnetic field with the one generated by another, independent LC circuit concatenated to the SQUID. A possible implementation of this idea is sketched in Fig.~\ref{fig:CircuitScheme}. In contrast to the open-waveguide experiment~\cite{Wilson-DCE-Analog-2011}, the opposite end of the CPW terminates here on a highly reflecting capacitive gap, so to obtain discrete high-Q cavity modes \cite{Johansson-PRA-2010,Wilson-PRL-2010}. This main goal of this section is to offer a quantitative estimate of the actual value of the opto-mechanical coupling that can be realistically obtained in state-of-the-art devices. 

The start point is the relation between the effective position $x_{\rm eff}$ of the analog mirror (measured from the physical position of the SQUID) and the magnetic flux $\phi$ threaded through the SQUID. Such a formula was derived in full detail in~\cite{Johansson-PRA-2010},
\begin{equation}
  x_{\rm eff}=\left(\frac{\Phi_0}{2\pi}\right)^2 \frac{1}{\ell_{wg}E_J(\phi)}
\end{equation}
where $\ell_{wg}$ is the impedance per unit length of the waveguide, $\Phi_0$ is the quantum of magnetic flux. Here, $E_J(\phi)$ is the (flux-dependent) Josephson energy of the SQUID, written as
\begin{equation}
  E_J(\phi)=2 E_J^o\,\left|\cos(\pi \phi/\Phi_0) \right|
  \end{equation}
in terms of the single junction Josephson energy $E_J^o$. Provided the modulation frequency is much smaller than the plasma
frequency of the SQUID $\omega_s=2\pi\sqrt{2E_J^o/\Phi_0^2C_J^o}$ (where $C_J^o$ is the capacitance of each Josephson junction forming the SQUID), a small time-dependent flux $\delta\phi$ then results in a time-dependent variation of the effective cavity length 
\begin{equation}
  \delta x_{\rm eff} = -x_{\rm eff} \frac{\delta E_J(\phi)}{E_J(\phi)}=x_{\rm eff} \frac{\sin(\pi \phi /\Phi_0)}{\cos(\pi \phi/\Phi_0)}\,\frac{\pi\,\delta\phi}{\Phi_0}.
  \label{eff_length}
\end{equation}

Assuming that the self-inductance of the SQUID is much smaller than the kinetic one, $L_{SQUID}\ll [\Phi_0/(2\pi)]^2/E_J^o$, the former can be neglected. The magnetic flux threaded by the $LC$ circuit through the SQUID can be written as $\delta\phi=M I_{LC}$ in terms of the current $I_{LC}$ flowing through the $LC$ and the mutual inductance $M$, this latter being of course bounded from above by the self-inductance of the circuit, $M/L_{LC}<1$.

Using the expression for the average magnetic energy stored in the ground state of the $LC$
\begin{equation}
  \frac{1}{2L_{LC}}\frac{[\varphi_{LC}^{(1)}]^2}{2}  = \frac{1}{4} \hbar \omega_{LC},
\end{equation}
in terms of the zero-point fluctuations $\varphi^{(1)}_{LC}$ of the magnetic flux, we can directly estimate $\varphi^{(1)}_{LC}$ in terms of circuit parameters, and then write the (operator-valued) magnetic flux threaded through the SQUID,
\begin{equation}
\hat{\delta \phi} = \frac{M}{ L_{LC}} \varphi^{(1)}_{LC}
\left(\frac{\hat{b}+\hat{b}^\dagger}{\sqrt{2}} \right)
\label{deltaPhi}
\end{equation}
in terms of the creation and destruction operators for the $LC$ harmonic oscillator, $\hat{b}$ and $\hat{b}^\dagger$ in our notation.

Inserting this expression into the one for the effective length \eqref{eff_length} and, this latter into the standard effective time-dependent Hamiltonian for the DCE emission in a cavity of average length $x^o$~\cite{Law-MirFieldInt-1995},
\begin{equation}
  H_{DCE}(t)= -\hbar \omega_a \frac{\delta x(t)}{2x^o} \left(\hat{a} +
\hat{a}^\dagger\right)^2
\end{equation}
and promoting the position $\delta x(t)$ to an operator, one gets to an effective coupling Hamiltonian between the $LC$ circuit and the (lowest) cavity mode in the desired form \eqref{Hint}, with a coupling constant
\begin{equation}
\hbar\omega_c=\frac{\hbar\omega_a}{4\sqrt{2}\pi}\frac{M}{L_{LC}}\frac{\omega_a}{I_J^o Z_{wg}}\sqrt{\hbar\omega_{LC} L_{LC}}\,\frac{\sin(\pi \phi /\Phi_0)}{\cos^2(\pi \phi /\Phi_0)}.
\label{omegaC_Iac}
\end{equation}
Here, we have considered the lowest mode of the waveguide with $\omega_a\approx \pi v/x^o$. Furthermore, $\omega_{LC}$ is the frequency of the LC circuit ($\omega_b$ in the rest of the article), $v=\sqrt{\ell_{wg}c_{wg}}$ and $Z_{wg}\equiv\sqrt{\ell_{wg}/c_{wg}}$ are respectively the velocity and the impedance of the waveguide mode in terms of the impedance $\ell_{wg}$ and capacitance $c_{wg}$ for unit length, $I^o_{J}=2\pi E_J^o/\Phi_0$ is the critical current of each Josephson junction forming the SQUID. A derivation of this same result starting from a more extended Lagrangian theory for the analogue system is reported in the appendix.

Plugging into this formula typical values for the SQUID device inspired from the experiment~\cite{Wilson-DCE-Analog-2011}, namely an operating frequency $\omega_a/(2\pi) \approx 5\,$GHz, an average cavity length of the order of a wavelength (in the waveguide) $x^o\approx 2\pi v/\omega_a$, a critical current $I^o_J\approx 1.25\,\mu$A, an impedance $Z_{wg}\approx 55\,\Omega$, an inductance $L_{LC}\approx 0.1\,$nH (of the order of the kinetic inductance of the Josephson junction), a flux concatenation ratio $M/L_{LC}=0.1$, and a trigonometric factor of order 1, one obtains a value for $\hbar \omega_c$ in the order of a few $10^4\,$Hz.
Given state-of-the-art values of the linewidths of superconductor-based oscillators in the tens of kHz range~\cite{Nori-CQED-Qfactor}, this value for $\hbar\omega_c$ is very promising in view of experimental observation of the dynamical Casimir-induced damping of the $LC$ circuit oscillations, as well as of the dynamical Casimir-induced periodic exchange of energy between the $LC$ circuit and the coplanar cavity. As it is shown in Fig.\ref{fig:CircuitScheme}, the $LC$ circuit is straightforwardly driven
and/or monitored just by coupling it to an external circuit: this provides the experimental access needed to implement both the free evolution and the driven-dissipative steady-state schemes discussed in the previous section. From a physical standpoint, the strong value of the analog opto-mechanical coupling can be understood in terms of the very light mass that the $LC$ circuit displays when viewed as an (analog) mechanical oscillator.

\section{Conclusions\label{sec:Conclusions}}

In this work we have theoretically studied a simplest system where the back-reaction effect of quantum fluctuations of the electromagnetic field onto a mechanically moving neutral object can be investigated. We have considered the simplest case of an optical cavity closed by a freely moving mirror attached to a spring. The mechanical motion of the mirror is responsible for the conversion of zero-point quantum fluctuations of the electromagnetic field into real photons via the dynamical Casimir effect, which can then be observed as propagating radiation. In return, the dynamical Casimir photons exert a friction force on the moving mirror that damps its motion. This quantum friction effect is studied in two most remarkable configurations. 

When no other external mechanical force is applied onto the mirror and the opto-mechanical coupling is relatively weak, the mirror motion performs periodic ring-down oscillations that are slowly damped out. The back-reaction appears as an additional contribution to the damping rate on top of standard friction. Since dynamical Casimir emission is strongest when the mechanical oscillations are on resonance with twice the cavity frequency, the two contributions to damping can be disentangled by looking at the variation of the mechanical damping rate as a function of the cavity frequency. As first predicted in\cite{Savasta-PRX-2018}, for strong values of the opto-mechanical coupling, the monotonic decay of the ring-down oscillations is replaced by a periodic exchange of energy between the mechanical and optical degrees of freedom in a sort of dynamical Casimir-induced two-photon Rabi oscillations.

When a periodically oscillating external force is applied to the mirror, the system is able to reach at long times a stationary state characterized by periodic oscillations of the mirror and a continuous emission of dynamical Casimir photons. In particular, we have shown how the properties of the back-reaction force can be extracted from the dependence of the mechanical oscillation amplitude on the frequency of the applied force. For weak opto-mechanical couplings, this response shows a single yet broadened peak whose linewidth carries an additional contribution from the back-reaction effect. For stronger couplings, the peak is replaced by a doublet whose splitting corresponds to the frequency of the periodic energy exchange between the mechanical and optical degrees of freedom.

Given the relatively small magnitude of the back-reaction effect in standard opto-mechanical devices based on macro- or mesoscopic mechanically moving mirrors, we have investigated its observability in analog models based on circuit-QED systems. Taking inspiration from the device recently used for the first observation of the dynamical Casimir effect~\cite{Wilson-DCE-Analog-2011}, we propose a configuration where the mechanically moving mirror is replaced by a LC circuit magnetically coupled to the SQUID that closes the co-planar waveguide into which the dynamical Casimir radiation is emitted. In such a system, the ring-down oscillations can be monitored by following in time the evolution of the oscillating current in the LC circuit. The response to the external force can be studied by sending an external monochromatic field onto the LC circuit and looking either at its current response or at the energy that is absorbed from the external field. The actual values of the system parameters that emerge from our simple modeling are extremely promising in view of experimental detection of the effect in state-of-the-art samples.

While the friction force of the dynamical Casimir effect onto the moving mirror is a simplest example of back-reaction effect of quantum fluctuations onto their environment, next theoretical steps will attack the far more difficult case of the back-reaction of Hawking radiation onto a black hole horizon. Schemes to study this physics in analog models based on condensed matter or optical systems are being explored, with special attention to unveiling analogies and differences with the late-time evaporation of astrophysical black holes.

\acknowledgements
We thank Andrea Vinante for helpful discussions on the experimental setups with superconducting circuits.
This work was supported by Julian Schwinger foundation, Grant No.~JSF-16-12-0001. I.C. acknowledges funding from Provincia Autonoma di Trento and from the EU-FET-Open grant MIR-BOSE Project No.737017.

\appendix*
\section{Lagrangian formulation of the LC-SQUID-CPW system}
\begin{figure*}[t]
\centering%
\includegraphics[width=0.8\linewidth]{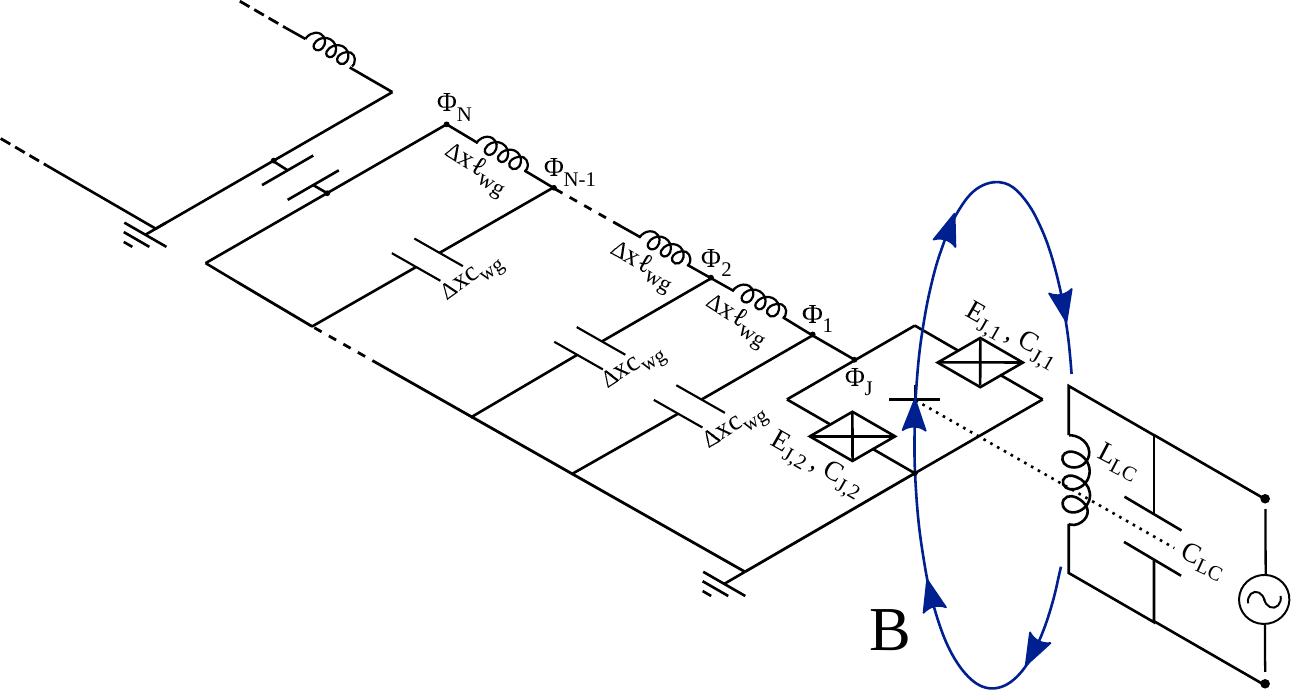}
\caption{Discrete model of the CPW magnetically coupled with the LC circuit.}
\label{fig:CircuitLump}
\end{figure*}
The following derivation is an extension of the Lagrangian formulation developed in \cite{Johansson-PRA-2010}, to the case in which the drive on the SQUID represents a dynamical degree-of-freedom for the system. Without affecting the generality of the following arguments, we assume the drive provided by a simple LC circuit that is magnetically coupled  to the SQUID. Other devices could have been considered to the same aim, such as another CPW, or any other electronic circuit that can be magnetically coupled to the SQUID.

For convenience, we start by writing the Lagrangian for the lumped-element model of the circuit depicted in Fig.~\ref{fig:CircuitLump}, and take the continuum limit after we calculate the equation of motion for the discrete degrees of freedom. Such a Lagrangian can be written as
\begin{equation}
	\mathcal{L}=\mathcal{L}_{\text{CPW}}+\mathcal{L}_\text{SQUID}+\mathcal{L}_\text{LC},
\label{Lcirc}
\end{equation}
where
\begin{align}
	\mathcal{L}_{\text{CPW}}&=\sum_{n=1}^{N-1}\left[\frac{1}{2}\Delta x \,c_{wg}\, \dot{\Phi}_n^2-\frac{1}{2}\frac{\left(\Phi_{n+1}-\Phi_n\right)^2}{\Delta x \,\ell_{wg}}\right],\label{Lcpw}\\
	\mathcal{L}_{\text{SQUID}}&=\sum_{j=1,2}\left[{\frac{1}{2}C_{J,j}\left(\dot{\Phi}_{J,j}\right)^2+E_{J,j}\cos\left(2\pi\frac{\Phi_{J,j}}{\Phi_0}\right)}\right],\label{Lsquid}\\
	\mathcal{L}_{\text{LC}}&=\frac{1}{2}C_{LC}\dot{\Phi}_{LC}^2-\frac{1}{2L_{LC}}\Phi_{LC}^2.\label{Llc}
\end{align}
are the Lagrangian for the CPW, the SQUID and the LC resonator respectively. Here we defined the flux quantum $\Phi_0\equiv \pi\hbar/e$ ($e$ is the electron charge), the capacitance $c_{wg}$ and inductance $\ell_{wg}$ densities in the CPW, the capacitance $C_{LC}$ and inductance $L_{LC}$ respectively for the capacitor and inductor in the LC resonator, as well as the capacitance $C_{J,j}$ and the Josephson energy $E_{J,j}=\hbar I_{c,j}/2e$ of the $j$th junction in the SQUID loop, characterized by the critical current $I_{c,j}$. We wrote the Lagrangian in Eqs.~(\ref{Lcpw}-\ref{Llc}) by assuming the node fluxes as generalized coordinates, which are defined as the time integral of the local voltage $V_j$
\begin{equation}
	\Phi_j(t)\equiv\int^t{d\tau V_j(\tau)}\,.
\label{Fluxes}
\end{equation}
Here $j=1,2,...,N$ denote the (discrete) degrees-of-freedom of the CPW, while $j=LC$ refers to the flux and the voltage across the LC resonator. We dropped from $\mathcal{L}_\text{SQUID}$ a term $\left(L I^2\right)_{SQUID}/2$, accounting for the magnetic energy stored in the SQUID because of the current $I_{SQUID}$ circulating in the loop. In other terms, we assumed the size of the SQUID loop small enough so that its self-inductance $L_{SQUID}$ is negligible compared to the Josephson inductances $L_{J,j}=\left(\Phi_0/2\pi\right)^2/E_{J,j}$. Given these assumptions, the fluxes $\Phi_{J,j}$ across the junctions can be directly related to the external flux piercing the loop as $\Phi_{J,1}-\Phi_{J,2}=\phi$, so that the SQUID can be described by the single degree of freedom $\Phi_J=\left(\Phi_{J,1}+\Phi_{J,2}\right)/2$. As a consequence, the SQUID behaves as a single Josephson junction described, in the simpler case of a perfectly symmetric junction characterized by the values $C_{J,1}=C_{J,2}=C_J^o/2$ and $E_{J,1}=E_{J,1}=E_J^o$, by the effective Lagrangian
\begin{equation}
	\mathcal{L}_{\text{SQUID}}=\frac{1}{2}C_J^o\dot{\Phi}_J^2+E_J\left(\phi\right)\cos\left(2\pi\frac{\Phi_J}{\Phi_0}\right).
\label{Lsquid2Sym}
\end{equation}
Here we indicated by $E_J\left(\phi\right)= 2 E_J^o\left|\cos\left(\pi\frac{\phi}{\Phi_0}\right)\right|$ the energy stored in the SQUID, which is a nonlinear function of the external flux piercing the flux. We work in the limit in which the plasma frequency $\omega_S^2=\left(2\pi/\Phi_0\right)^2\left(2E_J^o/C_J^o\right)$ of the SQUID far exceed the other characteristic frequencies in the circuit. In this regime, the oscillations of the phase across the SQUID are small, that is $\Phi_J/\Phi_0\ll 1$. Furthermore, we consider the external magnetic field piercing the SQUID to perform small oscillations around a bias value $\phi_b$, and we call $\delta\phi$ the amplitude of these oscillations, that are driven by the LC resonator magnetically coupled with its loop. With these assumptions we can approximate the SQUID Lagrangian by using the expansions
\begin{align}
	&\cos\left(2\pi\frac{\Phi_J}{\Phi_0}\right)\approx 1-\left(\frac{2\pi}{\Phi_0}\right)^2\frac{\Phi_J^2}{2},\label{SquidExp1}\\
	&E_J(\phi)=2E_J^o\left|\cos\left(\pi\frac{\phi}{\Phi_0}\right)\right|\nonumber\\
	&\hspace{14mm}=2E_J^o\left|\cos\left(\pi\frac{\phi_b+\delta \phi}{\Phi_0}\right)\right|\nonumber\\
	&\hspace{14mm}\approx 2E_J^o\cos\left(\pi\frac{\phi_b}{\Phi_0}\right)-2E_J^o\sin\left(\pi\frac{\phi_b}{\Phi_0}\right)\left(\pi\frac{\delta\phi}{\Phi_0}\right).
	\label{SquidExp2}
\end{align}
In the third line in Eq.~\eqref{SquidExp2} we assumed the amplitude of the oscillations $\delta\phi$ small enough so that the overall flux piercing the SQUID does not change sign. For the sake of brevity, we label in what follows $\varphi_b=\pi{\phi_b}/{\Phi_0}$, and write $\delta\phi=\chi\Phi_{LC}$, being $\chi\equiv M/L_{LC}$, where $M$ is the mutual inductance between the $LC$ and the SQUID and $\Phi_{LC}$ is the flux through the $LC$ circuit. Under these assumptions, the Lagrangian for the SQUID-LC subsystem can be written in the form
\begin{equation}
	\mathcal{L}_\text{SQUID}+\mathcal{L}_\text{LC}=\mathcal{L}'_{\text{SQUID}}+\mathcal{L}'_{\text{LC}}+\mathcal{L}_{\text{int}},
\label{Lsquid-lc}
\end{equation}
with
\begin{align}
	\mathcal{L}_{\text{SQUID}}'&=\frac{1}{2}C_J^o\dot{\Phi}_J^2-E_J^o\left(\frac{2\pi}{\Phi_0}\right)^2\cos\varphi_b \Phi_J^2,\label{Lsquid2}\\
	\mathcal{L}_{\text{LC}}'&=\frac{1}{2}C_{LC}\dot{\Phi}_{LC}^2-\frac{1}{2L_{LC}}\Phi_{LC}^2\nonumber\\
	&\hspace{20mm}-\chi E_J^o\left(\frac{2\pi}{\Phi_0}\right) \sin\varphi_b\Phi_{LC},\label{Llc2}\\
	\mathcal{L}'_{\text{int}}&=\chi\frac{E_J^o}{2}\left(\frac{2\pi}{\Phi_0}\right)^3\sin\varphi_b\Phi_J^2\Phi_{LC}.\label{Lsquid-lc2}
\end{align}
The Lagrangian in Eqs.~\eqref{Lsquid2} and \eqref{Llc2} describes the free evolution of the SQUID and the LC resonator respectively. There we notice the presence of a term linear in the flux $\Phi_{LC}$, which accounts for a shift of the equilibrium position of the oscillator, due to its coupling with the SQUID. The Lagrangian in Eq.~\eqref{Lsquid-lc2} is instead cubic in the products between the flux $\Phi_J$ across the junction and the flux $\Phi_{LC}$ across the inductance of the LC, and is responsible for a coupling between the two devices. In terms of Eqs.~\eqref{Lcpw} and (\ref{Lsquid2}-\ref{Lsquid-lc2}), the Lagrangian for the full circuit can thus be written as
\begin{equation}
	\mathcal{L}=\mathcal{L}_{\text{CPW}}+\mathcal{L}_{\text{SQUID}}'+\mathcal{L}_{\text{LC}}'+\mathcal{L}'_{\text{int}}.
\label{Lcirc2}
\end{equation}

\subsection{Equations of motion\label{subsec:CircuitEQMotion}}
\subsubsection{Radiation field\label{subsubsec:CircuitEQMotionRad}}
Basing on the effective Lagrangian in Eq.~\eqref{Lcirc2}, we determine here the equation of motion for the electromagnetic field. In the bulk region of the medium, in the continuum limit $\Delta x\rightarrow 0$, the field satisfies the wave equation
\begin{equation}
\frac{\partial^2\Phi}{\partial t^2}-v^2\frac{\partial^2\Phi}{\partial x^2}=0,
\label{WaveEq}
\end{equation}
where $v=1/\sqrt{c_{wg}\ell_{wg}}$ is the velocity of light in the CPW. Beside this, we need to pose opportune boundary conditions (BC) to the field. On the side opposite to the SQUID, that is at $x=-L$, such a BC is determined by the fact that the CPW is open and the current $I_{wg}$ need to be zero. Here the current is written in terms of the flux on the $N$ and $N-1$ node as $\left(\Phi_N-\Phi_{N-1}\right)=I_{wg}\left(\ell_{wg}\Delta x\right)$, from which follows in the continuum limit $I_{wg}=-\partial\Phi/(\ell_{wg}\partial x)$. This yields the first BC
\begin{equation}
\frac{\partial\Phi(t,-L)}{\partial x}=0.
\end{equation}
On the SQUID side instead, posing a BC means fixing the value of $\Phi(0,t)$, which corresponds to $\Phi_1$ in the discretized version of the Lagrangian in Eq.~\eqref{Lcpw}. It is important here to note that, in the model analyzed, $\Phi_1$ is not only a BC for the field, but it is a true dynamical quantity for the system. To determine the corresponding equation of motion, we notice that $\Phi_1$ coincides with the flux $\Phi_J$ across the junctions (see Fig.~\ref{fig:CircuitLump}). By posing $\Phi_J=\Phi_1$, and minimizing the Lagrangian in Eq.~\eqref{Lcirc2} with respect to variations in $\Phi_1$ we obtain, again in the continuum limit
\begin{multline}
C_J^o\ddot{\Phi}(t,0)+\frac{1}{\ell_{wg}}\frac{\partial\Phi}{\partial x}(t,0)+2E_J^o\left(\frac{2\pi}{\Phi_0}\right)^2\cos\varphi_b\,\Phi(t,0)\\-E_J^o \left(\frac{2\pi}{\Phi_0}\right)^3 \chi\sin\varphi_b\,\Phi(t,0)\Phi_{LC}=0.
\label{BCEqMot1}
\end{multline}
Since we work in the regime $\omega^2\ll\omega_S^2$, the first term in Eq.~\eqref{BCEqMot1} can be neglected, that reduces to
\begin{equation}
\Phi\left(t,0\right)+\frac{\partial\Phi}{\partial x}\left(t,0\right)\delta L_\eff=0.
\label{BCEqMot2}
\end{equation}
Here we defined the effective variation of the CPW length
\begin{equation}
	\delta L_\eff=\frac{1}{2E_J^o\ell_{wg}\cos\varphi_b}\left(\frac{\Phi_0}{2\pi}\right)^2\frac{1}{\left(1-\pi\chi\tan\varphi_b\Phi_{LC}/\Phi_0\right)}.
\end{equation}
To first order in $\Phi_{LC}/\Phi_0$, such an effective length is given by the sum of the two contributions
\begin{equation}
	\delta L_\eff=\delta L_\eff^{\phi_b}+\delta L_\eff^{\delta \phi}.
\label{Leff}
\end{equation}
Here
\begin{equation}
	\delta L_\eff^{\phi_b}\equiv\frac{1}{\cos\varphi_b}\frac{L_J}{\ell_{wg}}
\label{Leff2}
\end{equation}
is an effective length experienced by the CPW as an effect of the bias component $\phi_b$ of the magnetic flux concatenated with the SQUID, while
\begin{equation}
	\delta L_\eff^{\delta\phi}=\frac{\Phi_{LC}}{R}
\label{dLbias}
\end{equation}
with
\begin{equation}
	R= \left(\tan\varphi_b\pi\chi \delta L_\eff^{\phi_b}\right)^{-1}\Phi_0
\label{R}
\end{equation}
is an effective length induced by the drive. In Eq.~\eqref{Leff2} we introduced the characteristic inductance of the SQUID $L_J=\left(\Phi_0/2\pi\right)^2/(2E_J^o)$. For convenience we shift in what follows the origin of the $x$ coordinate by $L$, and rewrite the BCs obtained above as
\begin{align}
	&\frac{\partial\Phi(t,0)}{\partial x}=0,\label{BC1}\\
	&\Phi\left(t,L\right)+\frac{\partial\Phi}{\partial x}\left(t,L\right) \delta L_\eff=0.\label{BC2}
\end{align}
The former is satisfied by choosing field modes of the form $\cos(k_n x)$, while the latter sets the allowed values of the wavevector $\kappa_n$, that need to satisfy the following relation
\begin{equation}
	(\kappa_n \delta L_\eff)\tan(\kappa_n L)=1.
\label{BCkn}
\end{equation}
In the limit $\kappa_n \delta L_\eff\ll 1$, the BC at $x=L$ can be simplified as 
\begin{equation}
	\Phi(t,d)=0,
\label{BC2bis}
\end{equation}
with $d=L+\delta L_\eff$ the total effective length of the CPW. From the BC written in this form we find the allowed wavevectors $\kappa_n={(2n+1)\pi}/{2d(t)}$. The (normalized) basis functions, at the generic time instant $t$, can thus be written as 
\begin{equation}
\varphi_n(x)=\sqrt{\frac{2}{d(t)}}\cos(\kappa_n x),
\label{FieldExpansion}
\end{equation}
in terms of which the field in the CPW can be expanded as $\Phi(t,x)=\sum_n{Q_n(t)\varphi_n(x)}$, with $Q_n(t)$ the coefficients of the expansion, having the units $[flux]\times[length]^{1/2}$. Upon substitution of Eq.~\eqref{FieldExpansion} into the equation of motion in Eq.~\eqref{WaveEq}, we obtain the equation of motion for the $Q_n$
\begin{multline}
\ddot{Q}_n+\omega_n^2 Q_n-2\frac{\dot{d}}{d}\sum_k{\dot{Q}_n g_{nk}}\\-\left(\frac{\ddot{d}d-\dot{d}^2}{d^2}\right)\sum_k{Q_k g_{nk}}-\frac{\dot{d}^2}{d^2}\sum_{k,j}{Q_k g_{kj}g_{nj}}=0,
\label{EqMotCPW}
\end{multline}
with the coefficients
\begin{equation}
	g_{nk}=\begin{cases}
    \frac{(-1)^{n+k}}{2}\frac{(1+2k)(1+2n)}{k(k+1-n(n+1))}\phantom{000} \qquad\text{if}\, n\neq k,  \\
    0 \quad \phantom{\frac{(-1)^{n+k}}{2}\frac{(1+2k)(1+2n)}{k(k+1-n(n+1))}} \qquad  \text{if}\, n=k.
      \end{cases}
\end{equation}

\subsubsection{The LC resonator and its effective mass\label{subsubsec:CircuitEQMotionLC}}
In the previous section we derived the equation of motion for the field in the CPW. Since one of the BCs is non-stationary, we expanded the field in the instantaneous basis of eigenmodes $\{\varphi_n(x)\}$, and wrote the equation describing the time evolution for the coefficients $Q_k$ of such an expansion. This procedure is not new in literature, but has been  pursued in order to calculate the particle production from DCE or in cosmological scenarios as expanding universes for example []. What is different in the problem we study is that we consider the BC, that is  the $LC$ resonator in our case, as a truly dynamical object. In this section we study its dynamics, and derive the equation that describes the evolution in time of the effective length $d(t)$ of the CPW. The ultimate aim of this procedure is to introduce the effective mass for the BC, and provide an estimate for its value.

We start from the Euler-Lagrange equation for the LC resonator, that can be obtained directly from the Lagrangian in Eq.~\eqref{Lsquid-lc}. This has the form
\begin{multline}
\ddot{\Phi}_{LC}+\omega_{LC}^2\Phi_{LC}+\left(\frac{\Phi_0}{2\pi}\right)\frac{\chi}{2L_J C_{LC}}\sin\varphi_b\\
-\left(\frac{2\pi}{\Phi_0}\right)\frac{\chi}{4L_J C_{LC}} \sin\varphi_b\,\Phi^2(t,L)=0.
\label{phi_d}
\end{multline}
The value of the field $\Phi(t,L)$ at $x=L$ is obtained from the BC in Eq.~\eqref{BC2bis}
\begin{equation}
\begin{split}
\Phi(t,L)&=-\frac{\partial\phi}{\partial x}(d(t))(d(t)-L)\\
 &=\sqrt{\frac{2}{d}}\left(\sum_n{(-1)^n\,\kappa_n Q_n}\right)(d(t)-L),
 \label{pippo}
\end{split}
\end{equation}
where we used here the expansion $\Phi(t,x)=\sum_n{Q_n(t)\varphi_n(x)}$, along with the definition in Eq.~\eqref{FieldExpansion} for the field eigenmodes. Upon substitution of Eq.~\eqref{pippo} into Eq.~\eqref{phi_d} we can write the equation for the LC resonator as
\begin{multline}
\ddot{\Phi}_{LC}+\omega_{LC}^2\Phi_{LC}+\left(\frac{\Phi_0}{2\pi}\right)\frac{\chi}{2L_J C_{LC}}\sin\varphi_b\times\\
\left[1-\left(\frac{2\pi}{\Phi_0}\right)^2\frac{(d-L)^2}{d}\left(\sum_{n,k}{(-1)^{n+k}Q_n Q_k\kappa_n\kappa_k}\right)\right]=0.
\label{phi_d2}
\end{multline}
In writing Eq.~\eqref{phi_d2} we neglected a correction to the $LC$ frequency, induced by the  electromagnetic field in the CPW. In order to make connection with the optomechanical problem discussed in the previous sections, we write this equation in standard mechanical units and define an effective mass for the LC oscillator. To this aim we start from the free Lagrangian
\begin{equation}
 \mathcal{L}_{LC}=\frac{1}{2}C_{LC} \dot{\Phi}_{LC}^2-\frac{1}{2L_{LC}}\Phi_{LC}^2,
\end{equation}
and write it in terms of the effective length $d$ defined above. By using Eq.~\eqref{dLbias}, this takes the form
\begin{equation}
 \mathcal{L}_{\text{LC}}=\frac{1}{2}C_{LC} R^2\dot{d}^2-\frac{1}{2L_{LC}}R^2 \left[d-(L+\delta L_\eff^{\phi_b})\right]^2.
\end{equation}
The momentum conjugate to the effective length $d$ is 
\begin{equation}
 p=\frac{\partial\mathcal{L}_{\text{LC}}}{\partial \dot{d}}=C_{LC} R^2\dot{d},
\end{equation}
and allows us to identify the effective mass $m=R^2 C_{LC}$. Considering typical values for the physical parameters of the system, such an effective mass can take values of the order
\begin{equation}
\begin{split}
	m&=C_{LC} R^2=C_{LC}\left(\frac{\cos\varphi_b}{\tan\varphi_b}\frac{\ell_{wg}}{L_J}\frac{\Phi_0}{\pi\chi}\right)^2\\
	  &\sim 10^{-30}\,\text{Kg}.
\end{split}
\label{EffMass}
\end{equation}
In terms of these quantities, the Eq.~\eqref{phi_d2} can be rewritten as
\begin{equation}
m\,\ddot{d}+m\omega_{LC}^2\left(d-L_{\text{eq}}\right)
-\frac{1}{d}\sum_{kn}{(-1)^{n+k}\tilde{Q}_n\tilde{Q}_k\omega_n\omega_k}=0.
\label{EqMotLC4}
\end{equation}
Here we defined the quantities
\begin{align}
	\tilde{Q}_n&\equiv\left(\frac{2\pi}{\Phi_0}\right)\delta L_\eff^{\phi_b}\left(\frac{m a}{v^2}\right)^{1/2}{Q}_n\\
	L_{\text{eq}}&\equiv L+\delta L_\eff^{\phi_b}-\frac{a}{\omega_{LC}^2}\\
	a&\equiv\frac{R\Phi_0}{2\pi}\frac{\chi}{2L_Jm}\sin\varphi_b
\end{align}
and used the zeroth order approximation
\begin{equation}
\frac{\left(d-L\right)^2}{d}\approx\frac{\left(\delta L_\eff^{\phi_b}\right)^2}{d}.
\label{expans}
\end{equation}
The coefficients $\tilde{Q}_n$ have here the units $[length]\times[mass]^{1/2}$, and the Eq.~\eqref{EqMotLC4} is equivalent to the equation of motion of a mirror interacting with an electromagnetic field via its radiation pressure. We do not go through the quantization procedure for this theory. It is laborious and already addressed in [Law]. For our purposes it is sufficient to remember that, in the regime of small oscillations of the mirror around its equilibrium position, the quantized theory leads to the Hamilintonian we used in Eq.~\eqref{Hint} to describe the interaction. By taking advantage of this analogy, the value of the effective coupling constant can be calculated by using the definition in Eq.~\eqref{omegaC} given in Sec.~\ref{sec:System} (with $\omega_{LC}$ in place of $\omega_b$). Given the expression for the effective mass in Eq.~\eqref{EffMass}, this reproduces the result in Eq.~\eqref{omegaC_Iac} of the main text.


\bibliography{DCE-BR.bib}
\bibliographystyle{unsrt}

\end{document}